\documentclass[aps,11pt,superscriptaddress,nofootinbib,notitlepage,showkeys]{revtex4-2}
\pdfoutput=1

\usepackage{comment}
\usepackage{array}
\usepackage{dcolumn}
\usepackage{amsmath,amssymb,amsfonts,graphicx,comment,bm,mathtools}
\usepackage[normalem]{ulem}
\usepackage[dvipsnames]{xcolor}
\usepackage{url}
\usepackage{tikz}
\usepackage{grffile}

\usepackage{hyperref}
\usepackage[english]{babel}
\usepackage{hyperref}
\usepackage{epstopdf, epsfig}
\usepackage{natbib}
\usepackage{amsthm}
\usepackage{todonotes}
\usepackage{color}
\makeatletter\AtBeginDocument{\let\@elt\relax}\makeatother

\newcommand{\mE}{{\mathcal  E}}
\newcommand{\mN}{{\mathcal  N}}
\newcommand{\mW}{{\mathcal  W}}

\newcommand{\bk}{{\bf k}}
\newcommand{\de}{{\text{d}}}

\newcommand{\be}{\begin{equation}}
\newcommand{\ee}{\end{equation}}

\newcommand{\si}{\color{purple}}

\begin{document}
\title{Dynamical cooling driven by self-similar fronts
in  the 2D nonlinear Schrödinger model}

\author{Jason Laurie}
\affiliation{Department of Mechatronics and Biomedical Engineering, College of Engineering and Physical Sciences, Aston University, Birmingham, B4 7ET, United Kingdom}

\author{Simon Thalabard}
\email{Simon.Thalabard@univ-cotedazur.fr}
\affiliation{Institut de Physique de Nice, Universit\'e C\^ote d'Azur CNRS - UMR 7010, 17 rue Julien Laupr\^etre, 06200 Nice, France}

\author{Sergey Nazarenko}
\email{Sergey.Nazarenko@univ-cotedazur.fr}
\affiliation{Institut de Physique de Nice, Universit\'e C\^ote d'Azur CNRS - UMR 7010, 17 rue Julien Laupr\^etre, 06200 Nice, France}

\graphicspath{{figs/new/}}
\date{\today}

\begin{abstract}
\noindent We analyze the dynamics towards partial thermalization and subsequent cooling in the defocusing two-dimensional nonlinear Schrödinger model, using direct simulations and insights from the wave-kinetic equations (WKE) and a fourth-order differential approximation model (DAM).   We show that the evolving WKE spectrum exhibits two  distinct similarity ranges--the quasi-thermal core and the ultraviolet tail--whereas in the DAM, an additional range of infrared self-similarity appears. 
By  stretching  the quasi-thermal region, the self-similar fronts drive  an effective dynamical cooling process towards the formal but ill-defined equilibrium state at vanishing temperature---analogous to an ultraviolet catastrophe
in a system of classical waves. 
\end{abstract}

\maketitle
\tableofcontents
\section{Introduction}\label{sec:intro}

\noindent The two-dimensional (2D) nonlinear Schrödinger (NLS) equation is a fundamental partial differential equation in theoretical physics, arising in various physical contexts, from nonlinear optics and water-wave theory to Bose-Einstein condensations. It is an evolution equation for the complex wave function $\psi({\bf x},t)$ which is, in non-dimensional form, given by 
\begin{align}\label{eq:2d_nls}
i\frac{\partial \psi}{\partial t} = - \nabla^2\psi + |\psi|^2 \psi.
\end{align}
Depending on the context, $|\psi|^2$ can represent the intensity of light propagating through a nonlinear Kerr medium, the envelope intensity of a surface wave train, or the number of particles in a Bose-Einstein condensate.

In a 2D doubly-periodic domain $[0,L)\times[0,L)$, we can represent the physical-space wave function in terms of the Fourier series, $\psi({\bf x}, t) = \sum_{\bf k}\, \hat{\psi}_{\bf k}(t)\, e^{i{\bf k}\cdot{\bf x}}$, where $\hat{\psi}_{\bf k}(t)=\hat{\psi}({\bf k},t) = (1/L^2)\int \psi({\bf x},t)\,e^{-i{\bf k}\cdot{\bf x}}\, \de^2{\bf x}$ are the respective Fourier amplitudes. The machinery of wave turbulence~\cite{zakharov_kolmogorov_1992,nazarenko_wave_2011} eventually provides a kinetic description of the 2D NLS~\eqref{eq:2d_nls} under a certain number of specific assumptions: (i) infinite box limit $L\to \infty$, (ii) small nonlinearity and (iii) random phase and amplitudes.  The resulting evolution is described by the following wave-kinetic equation (WKE), 
\begin{align}\label{eq:kinetic_omega}
& \frac{\partial N_\omega}{\partial t} =
\!  \int\! \!S_{\omega,\omega_1}^{\omega_2,\omega_3}N_\omega N_{\omega_1}N_{\omega_2} N_{\omega_3}\left(\frac{1}{N_\omega}+\frac{1}{N_{\omega_1}}-\frac{1}{N_{\omega_2}}-\frac{1}{N_{\omega_3}}\right) 
\delta({\omega+\omega_1-\omega_2-\omega_3})\, \de\omega_1\,\de\omega_2\,\de\omega_3,
\end{align}
where the spectrum is assumed to be isotropic and described by the frequency space wave action density defined as
\begin{align}
N_\omega(t)  = N(\omega,t) = \dfrac{1}{2} \int n({\bf k},t) \delta(|\bk|^2-\omega) \ \de^2\bk,\quad \text{with}\quad   n({\bf k},t)= \lim_{L\to \infty} \left(\frac{L}{2\pi}\right)^2 \left\langle |\hat{\psi}_{\bf k}(t)|^2\right\rangle.
	\label{eq:N}
\end{align}
Here, the angle brackets $\langle \cdot \rangle$ denote averaging over different ensembles of initial conditions, and  $\omega(\bk)=|\bk|^2$ is the wave frequency. 
 Equation \eqref{eq:kinetic_omega} features the kernel $S_{\omega,\omega_1}^{\omega_2,\omega_3}=S({\omega},{\omega}_1,{\omega}_2,{\omega}_3)$ which emits a degree of homogeneity of $-1$ in frequencies:  $S(\lambda{\omega},\lambda{\omega}_1,\lambda{\omega}_2,\lambda{\omega}_3) = \lambda^{-1}S({\omega},{\omega}_1,{\omega}_2,{\omega}_3)$ \cite{dyachenko_optical_1992}---see Appendix \ref{app:convergence} for its explicit  expression.

The  WKE ~\eqref{eq:kinetic_omega}  conserves the total energy  and the total wave action (number of particles) per unit area,\emph{e.g.} 
\begin{align}
\label{eq:EN}
\mE=\int_{\mathbb R^+}\omega N_\omega \, \de\omega \quad \hbox{ and  } \quad \mN=\int_{\mathbb R^+} N_\omega \,\de\omega.
\end{align}
The general thermal equilibrium solution of the WKE \eqref{eq:kinetic_omega} is known as the Rayleigh-Jeans (RJ) wave action spectrum
\begin{align}\label{eq:Rayleigh-Jeans}
N^{RJ}_\omega(T,\mu) = \frac{T}{\mu + \omega},
\end{align} 
where $T$ and $\mu$ are two positive parameters, which can be interpreted as the temperature and chemical potential, respectively. 
It is clear that for any $T> 0$, the energy and the particle integrals~\eqref{eq:EN} diverge. Therefore, for any initial state with finite  densities of energy and wave action, the ultimate thermal equilibrium could only form with $T=0$. This is the  statement of the famous ``ultraviolet catastrophe" formulated for classical waves.
Despite the fact that the equilibrium state  at vanishing temperature    is trivial, the evolution of the system toward equilibrium raises interesting physical questions. One could want to picture this evolution as a cooling process, but the system is necessarily  out of equilibrium during this stage. Hence, it is not \textit{a priori} clear that it can be meaningfully (or unambiguously) characterized by any time-dependent temperature or chemical potential. More broadly, one could ask if the cooling process exhibit  any universal structure or scaling laws.  In the present work, we aim to address these questions by studying the case of the 2D NLS system. 

Our approach is based on the identification of self-similar solutions to describe the evolution of the various portions of the profile.
To guide our intuition, we will discuss our results alongside the analysis of a fourth-order diffusion approximation to the WKE, the so-called differential approximation model (DAM)~\cite{dyachenko_optical_1992, nazarenko_wave_2011, thalabard2021inverse}, which replaces the collision integral of the WKE~\eqref{eq:kinetic_omega} by a fourth-order differential equation, while keeping the scaling properties and the equilibrium solutions of the collision integral.
The convergence towards self-similar regimes is assessed through numerical simulations of both the 2D NLS and the DAM systems.

The paper is structured as follows \S\ref{sec:2D} presents the DAM and its analogy to the 2D NLS, with \S\ref{sec:setups} outlining the numerical setups. \S\ref{sec:kinematics} describes the kinematics of relaxation in terms of quasi-thermalized RJ profiles before we examine the self-similar characteristics at the high-frequency end of the evolving spectrum in \S\ref{sec:SSHF}. In \S\ref{sec:SSDAM} we discuss the relaxation in the DAM from the point of view of self-similar solutions, and extend the discussion to the NLS in \S\ref{sec:SSNLS}. \S\ref{sec:concluding} formulates concluding remarks.

\section{2D  kinetics under super-local interactions}
\label{sec:2D}
Upon assuming that the main contributions to the collision integral 
forming the right-hand side of the WKE~\eqref{eq:kinetic_omega} occur when all wave frequencies are close in magnitude, one can dramatically simplify the kinetic wave description into one of an ordinary differential equation. This assumption is known as the super-local wave interaction hypothesis. 
Mathematically, one considers that the main contributions to the WKE~\eqref{eq:kinetic_omega}  occur when $\omega_i \approx \omega(1+p_i)$ where $|p_i|\ll 1$ for $i=1,2,3$, so that Taylor expansions in $p_i$ can be made.  This ad-hoc strategy leads to the fourth-order local approximation, which reduces the multi-dimensional integration to localized operators in the form of a differential approximation model (DAM)~\cite{dyachenko_optical_1992}
\begin{align}\label{eq:dam}
  \frac{\partial N(\omega,t)}{\partial t} =  S_0\frac{\partial^2}{\partial \omega^2}\left( \omega^5 N^4 \frac{\partial^2}{\partial \omega^2}\left( \frac{1}{N}\right)\right),
\end{align}
where the parameter $S_0$ is thereafter assumed to be unity (which amounts to rescaling time).
Equation~\eqref{eq:dam}  represents  a nonlinear continuity equation  for the wave action density $ N_\omega = N(\omega, t)$ by observing
\begin{align}
\label{eq:nDAM}
	\frac{\partial N_\omega }{\partial t} + \frac{\partial Q}{\partial \omega}=0, \qquad  Q = -\frac{\partial K }{\partial \omega}, \qquad K= S_0\omega^{5} N_\omega^{4} \frac{\partial^2}{\partial {\omega^2}} \left(\frac{1}{N_\omega}\right).
\end{align}
The quantity $Q$ represents the flux of wave action under the super-local approximation. As the original WKE conserves both the wave action and quadratic energy per unit area (as defined in \eqref{eq:EN}), the DAM can also be recast as a continuity equation for the quadratic energy density $E_\omega = E(\omega,t) = \omega N_\omega$,  involving the quadratic energy flux $P$:
\begin{align}
\label{eq:eDAM}
	\frac{\partial E_\omega}{\partial t}  + \frac{\partial P}{\partial \omega}=0, \quad P = \omega Q+K.
\end{align}
Also, the DAM has the same family of equilibrium RJ solutions as the WKE given by~\eqref{eq:Rayleigh-Jeans}, which can form with $T>0$ only if the frequency space is bounded from above. However, as we will see below, the finite temperature RJ spectra appear, in a truncated form, as  transient states during the evolution.

\section{Numerical setups}
\label{sec:setups}
\subsection{2D Nonlinear Schrödinger Equation}

\noindent We perform an ensemble of four direct numerical simulations of the 2D NLS equation~\eqref{eq:2d_nls} using a pseudospectral spatial discretization of resolution $2048\times 2048$ uniformly spaced grid points in a square periodic domain of size $L=2\pi$. The pseudospectral method is fully dealiased with the $3/2$-rule on the nonlinear term. Equation~\eqref{eq:2d_nls} is time-integrated using a fourth-order Runge-Kutta exponential time-differencing scheme~\cite{kassam_fourth-order_2005,cox_exponential_2002} with a fixed timestep of $\Delta t=1\times 10^{-6}$. High frequency dissipation is added to the right-hand side of Eq.~\eqref{eq:2d_nls} in the form of hyperviscosity 
$-i\nu(-\nabla^2)^{8}\psi({\bf x},t)$  with $\nu = 5\times10^{-47}$. This is to ensure that we inhibit the formation of an artificial high-frequency bottleneck at late times due to the presence of a maximum frequency $\omega_{\rm max} =N^2/4$. Each ensemble is initiated with a randomized initial condition such that the initial wave function has the form
\begin{align*}
	\psi({\bf x},0)&=\sum_{\bf k}  \hat{\psi}_{\bf k}(0)e^{i{\bf k}\cdot{\bf x} + i\theta_{\bf k}}\quad
   \text{with}\quad \hat{\psi}_{\bf k}(0) = A e^{-\ln^2(k/k_{0})/2\sigma_{0}^2},
\end{align*}
peaked at frequency $\omega_{0} = k_{0}^2 = 16^2 =  256$ with $\sigma_{0}=0.05$ with amplitude $A=0.3162$. Here, $\theta_{\bf k}$ represents a uniformly distributed variable $\theta_{\bf k}\in [0,2\pi)$ independently sampled for each wave vector ${\bf k}$. Subsequently, each ensemble leads to an initial identical wave action spectrum profile but involving independent random phases.  The system initially evolves so that the total wave action $\mathcal{N}$ and the total energy $\mathcal{E}$ are conserved until the wave action spectrum reaches the dissipation located at high frequency at late times. 

\subsection{Differential Approximation Model}

\noindent For the DAM, numerical integration is performed using the two-step Adams–Bashforth method in time and the central difference scheme in  frequency space. We start from the initial condition 
\begin{align}
	 N_\omega(t=0)=  e^{-(\omega-\omega_0)^2/2\sigma_0^2}\quad \text{with  $\sigma_0=0.1$, $\omega_0  =10^{13}$},
\end{align} 
in a domain logarithmically discretized between $\omega_{\min} =8.7\times 10^{-19}$ and $\omega_{\max}=1.1 \times 10^{18}$ using 2,400 collocation points. No dissipation is used. 

\section{The kinematics of cooling}
\label{sec:kinematics}

\begin{figure}
\includegraphics[width=0.49\textwidth]{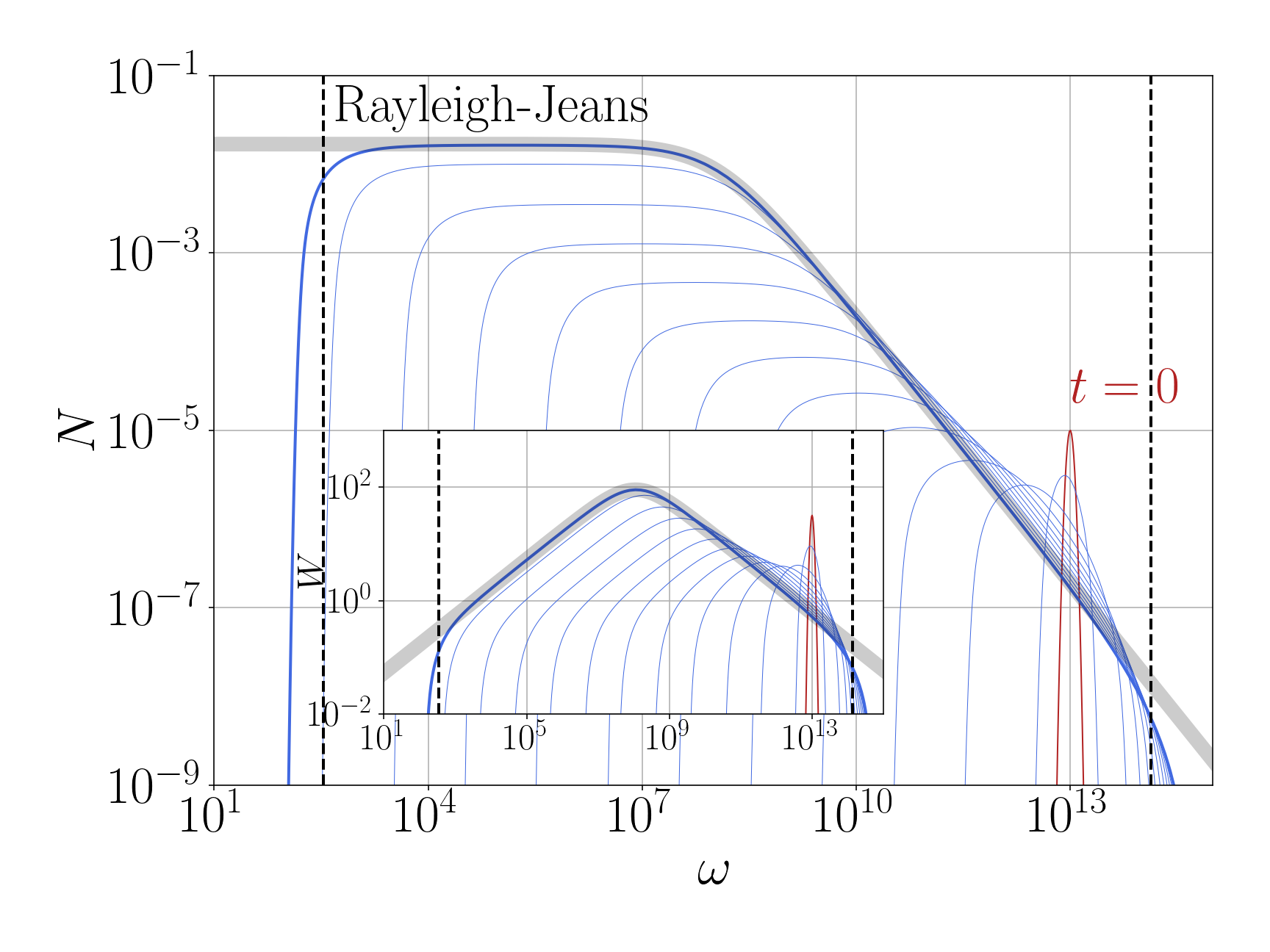}
\includegraphics[width=0.49\textwidth]{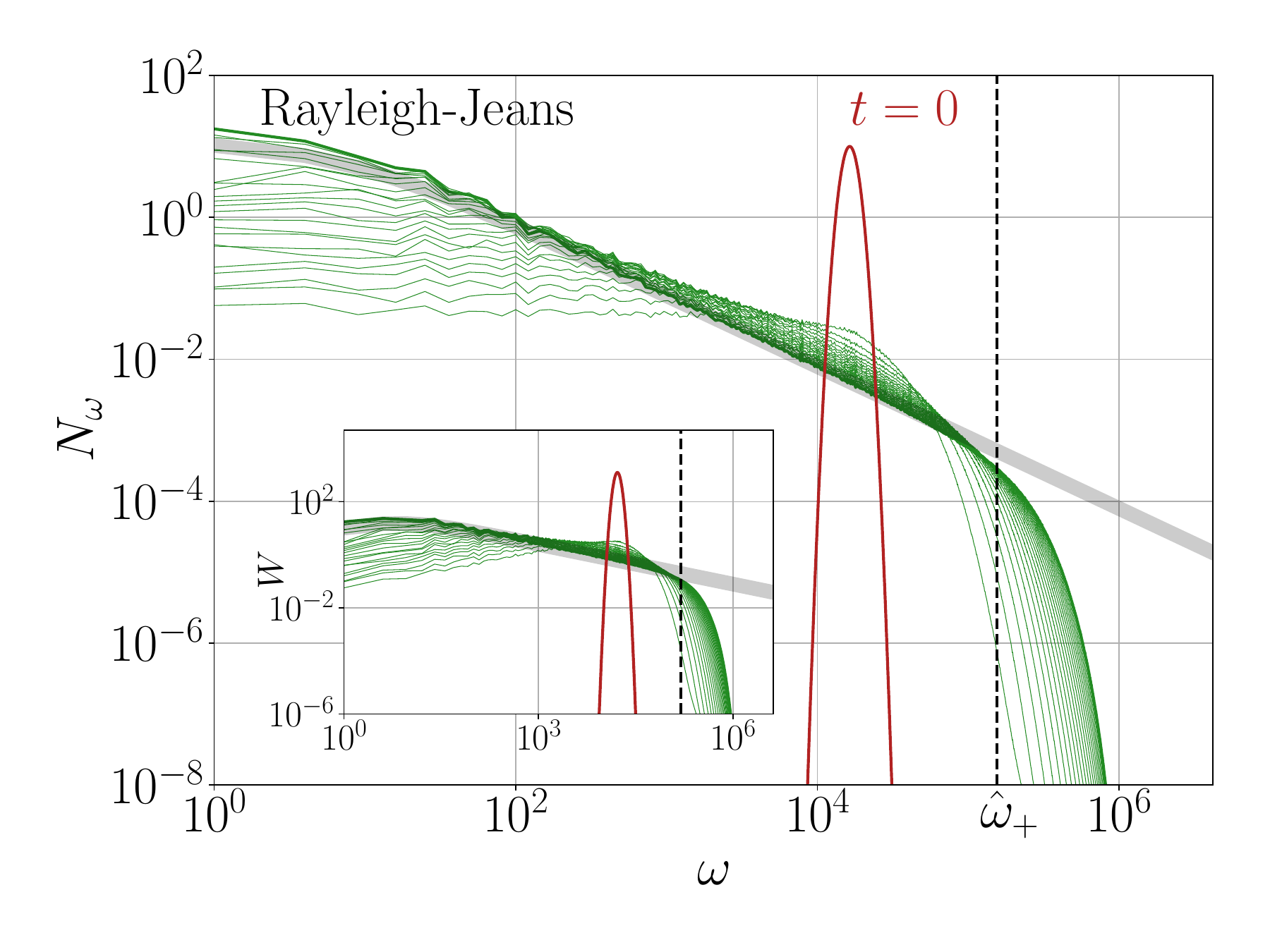}
\caption{(Left: DAM and Right: NLS) Temporal evolution of the wave action spectrum $N(\omega,t)$  towards the Rayleigh-Jean spectrum (thick gray curve). The initial spectra at $t=0$ is given by the red curve. The vertical black dashed lines indicate estimates of the left-front $\hat{\omega}_-$ (DAM only) and the right-front $\hat{\omega}_+$ at the final time. Insets show the compensated spectrum $W(\omega,t) := \omega^{1/2} N(\omega,t)$.}
\label{fig:1}
\end{figure}

\noindent First of all, let us make a general observation arising from the numerical simulations. In both DAM and NLS, the free evolution of an initial condition with a narrow-band spectrum appears to be well-approximated by the quasi-equilibrium ansatz prescribed by a time-dependent RJ profile~\eqref{eq:Rayleigh-Jeans}, namely $N_\omega(t) \approx N^{RJ}_\omega(\mu(t),T(t))$, over a finite but increasing range of frequencies; see Fig.~\ref{fig:1}. For the DAM, this  quasi-thermal  range has both lower and upper boundaries $\omega \in (\hat{\omega}_-,\hat{\omega}_+)$. This comes from the  interactions in DAM being super-local, resulting in  initially localized spectrum  retaining finite support in $\omega$ (at least over sufficiently short times).
Discrepancies with RJ occur across the  (ultraviolet) frequencies: $\hat{\omega}_+ <\omega <\omega_+$, and the (infrared) range 
$\omega_-<\omega<\hat{\omega}_- $,  where $\omega_-$ and $\omega_+$ are  the positions of the left and right fronts of the spectrum, respectively. Both are well-defined for the DAM and  practically estimated by thresholding the  propagating front to a very small value.
On the other hand, for NLS $\hat{\omega}_-=0$. This can be understood as a result of interaction nonlocality in the WKE for sufficiently flat spectra, which dictates that the WKE solution near zero frequency must be approximately independent of $\omega$  (see Appendix~\ref{app:nonlocal}).  Although it is in principle well-defined, the right front $\omega_+$ proves hard to track numerically due to numerical noise and less extended frequency ranges. Besides its properties are non-universal because of the presence of dissipation term. We therefore restrict our analysis to $\hat \omega_+$ for NLS.

From  quasi-thermal RJ range,  one can define time-dependent  (effective)  temperature and potential to parameterize  the spectral kinematics. The dynamics  can be interpreted as an effective dynamical cooling for the quasi-thermal core. However, the values of  temperature and potential are not fully universal and depend on their precise fitting schemes. 
On the one hand, a natural fitting scheme for $T$ and $\mu$ is obtained by monitoring the value and frequency of 
\begin{align}
	\label{eq:W}
	W_\infty(t) := \sup_{\omega} W(\omega,t),\quad W(\omega,t) := \omega^{1/2} N_\omega,
\end{align}
namely the maximum of the symmetric profile $W(\omega,t) := \omega^{1/2} N_\omega$, see insets in Fig.\ref{fig:1}. Under the approximation
$N_\omega\simeq N^{RJ}_\omega$, the temperature $T$ and the chemical potential $\mu$ are then given by the formulae 
\begin{align} \label{eq:muT}
\mu(t) = \underset{\omega}{\operatorname{argmax}} \ W(\omega,t),\quad \text{and}\quad T(t)= 2 \mu^{1/2} W_\infty(t).
\end{align}

On the other hand, the joint conservation of wave action $\mN=N_0$ and energy $\mE=E_0$ suggests another fitting scheme, which  relates the long-time asymptotics of the chemical potential and temperature to the right-front dynamics. To distinguish these  from the thermodynamic estimates \eqref{eq:muT}, we denote them by $\hat{\mu}$ and $\hat{T}$.  Let us assume that the RJ spectrum is realized in the range $\omega <\hat{\omega}_+$, and that the amount of energy and particles are negligible at $\omega >\hat{\omega}_+$. 
Note that for  $\hat{\mu}  \gg \hat{\omega}_-$ the amount of energy and particles at $\omega \lesssim  \hat{\omega}_-$ are also negligible. Then for $\hat{\mu}  \ll \hat{\omega}_+$ we have
\begin{align}
\label{eq:EN1}
& \mE=\int_{0}^{\hat{\omega}_+}\omega N^{RJ}_\omega \, \de\omega \approx \hat T  \hat{\omega}_+ \quad \text{and} \quad \mN=\int_{0}^{\hat{\omega}_+} N^{RJ}_\omega \,\de\omega \approx \hat T  \ln \left(\frac{\hat{\omega}_+}{\hat \mu}\right).
\end{align}
This gives 
\begin{align}\label{eq:muT-cons}
    \hat T= \frac{E_0}{\hat{\omega}_+},\quad \hat{\mu}= \hat{\omega}_+ e^{-\frac{\hat{\omega}_+}{\omega_0}},\quad  \text{where}\quad \omega_0 = \dfrac{E_0}{N_0}.
\end{align}
\begin{figure}
\includegraphics[width=\textwidth]{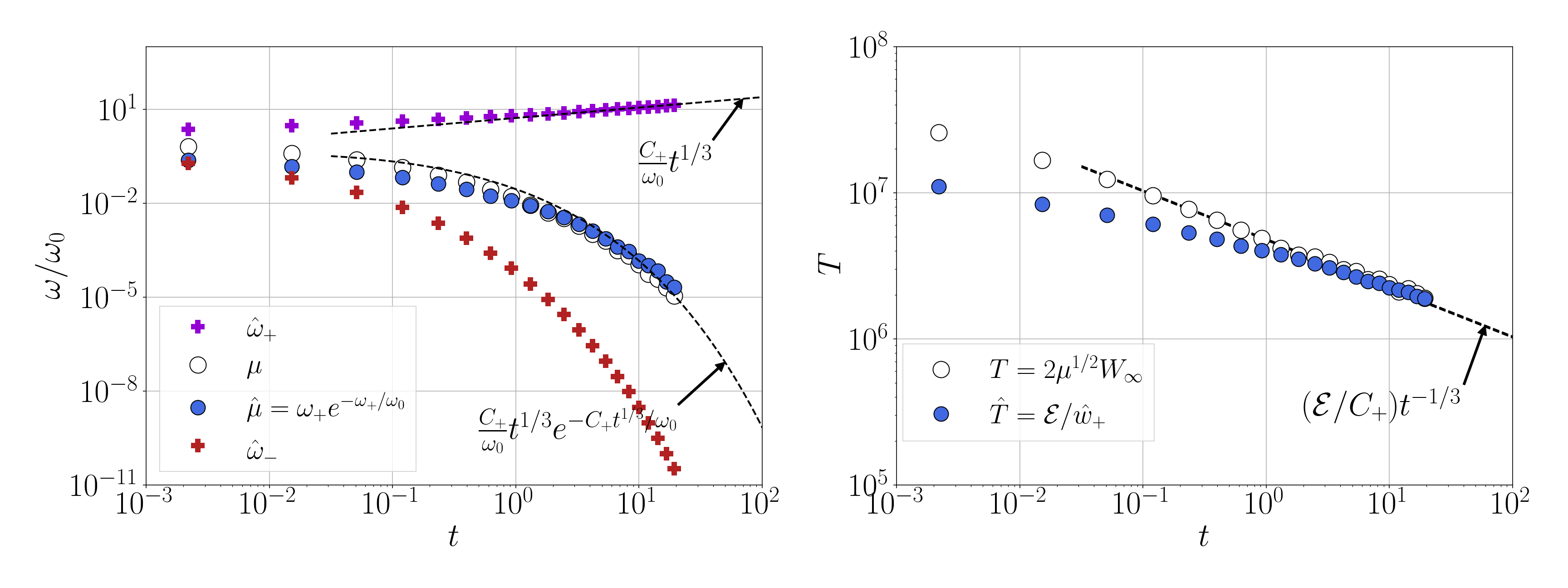}\\
\includegraphics[width=\textwidth]{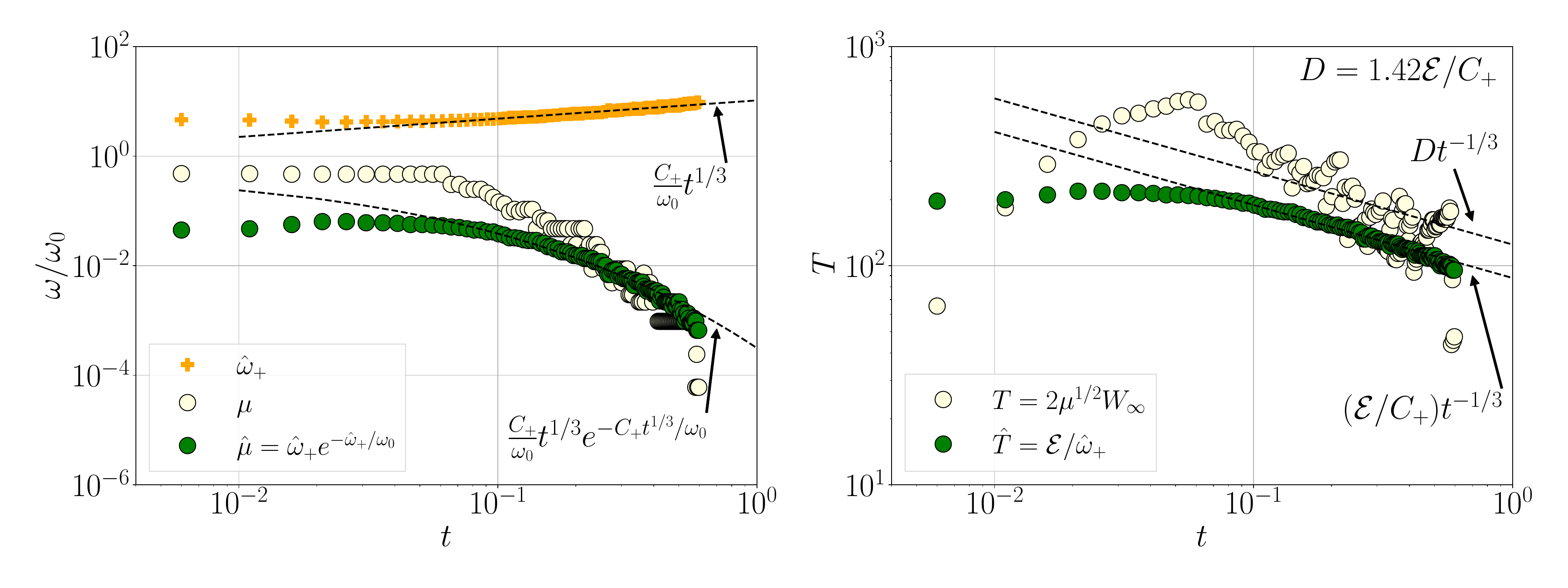}
\caption{
(Top: DAM and Bottom: NLS). Left: Temporal dynamics of the effective Rayleigh-Jeans chemical potentials $\mu$ and $\hat{\mu}$, together with the estimates of the propagating front frequencies $\hat{\omega}_-$ (only for DAM) and $\hat{\omega}_+$,  in units of $\omega_0 = E_0/N_0$. Right: Temporal dynamics of the effective Rayleigh-Jeans temperatures $T$ and $\hat T$. We estimate $\hat{\omega}_+$ by the frequency at which the energy spectrum drops to $0.4$ for DAM and $0.7$ for NLS of its maximum.
\label{fig:2}
}
\end{figure}
\begin{figure}
\includegraphics[width=0.49\textwidth]{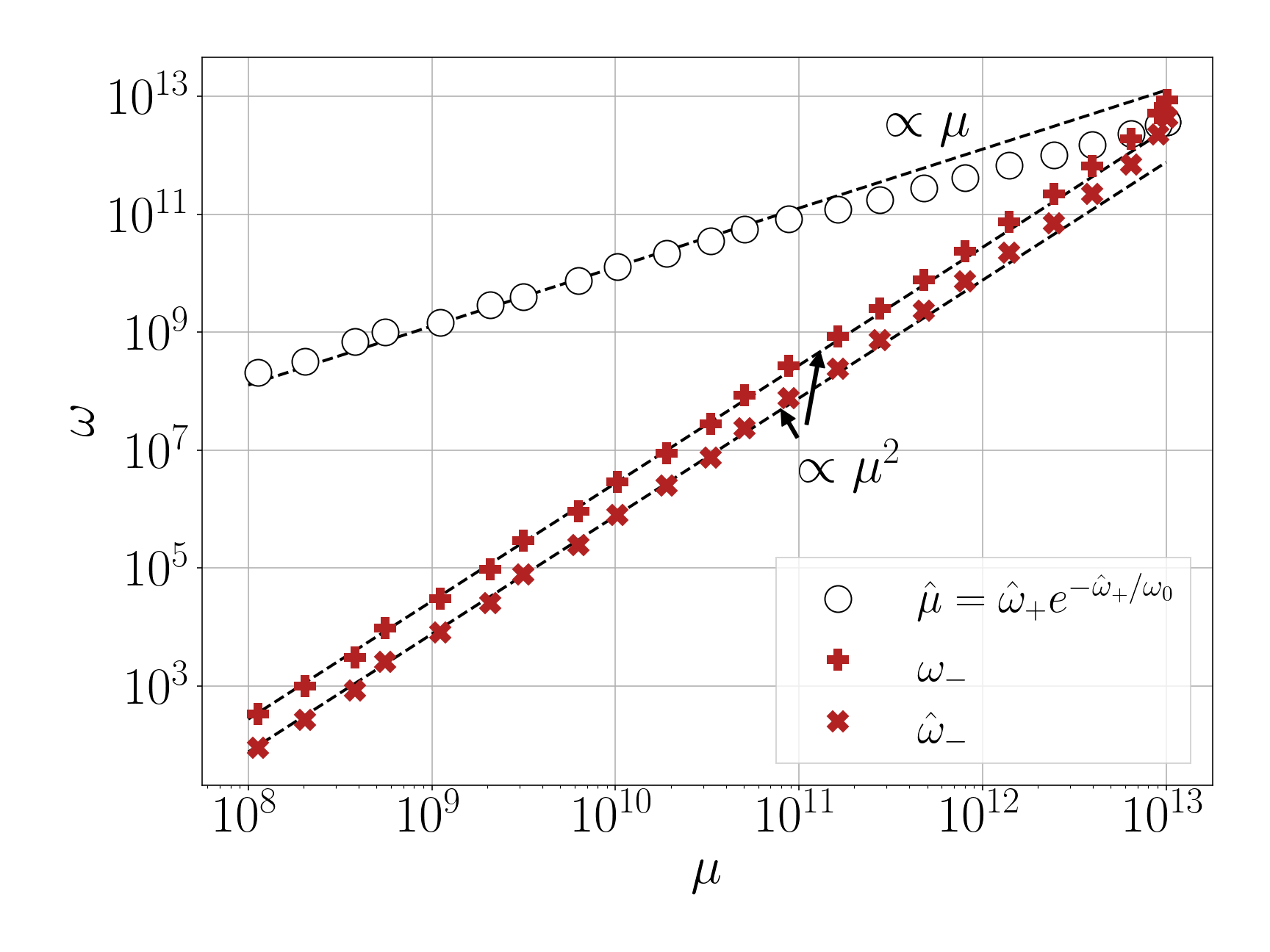}
\includegraphics[width=0.49\textwidth]{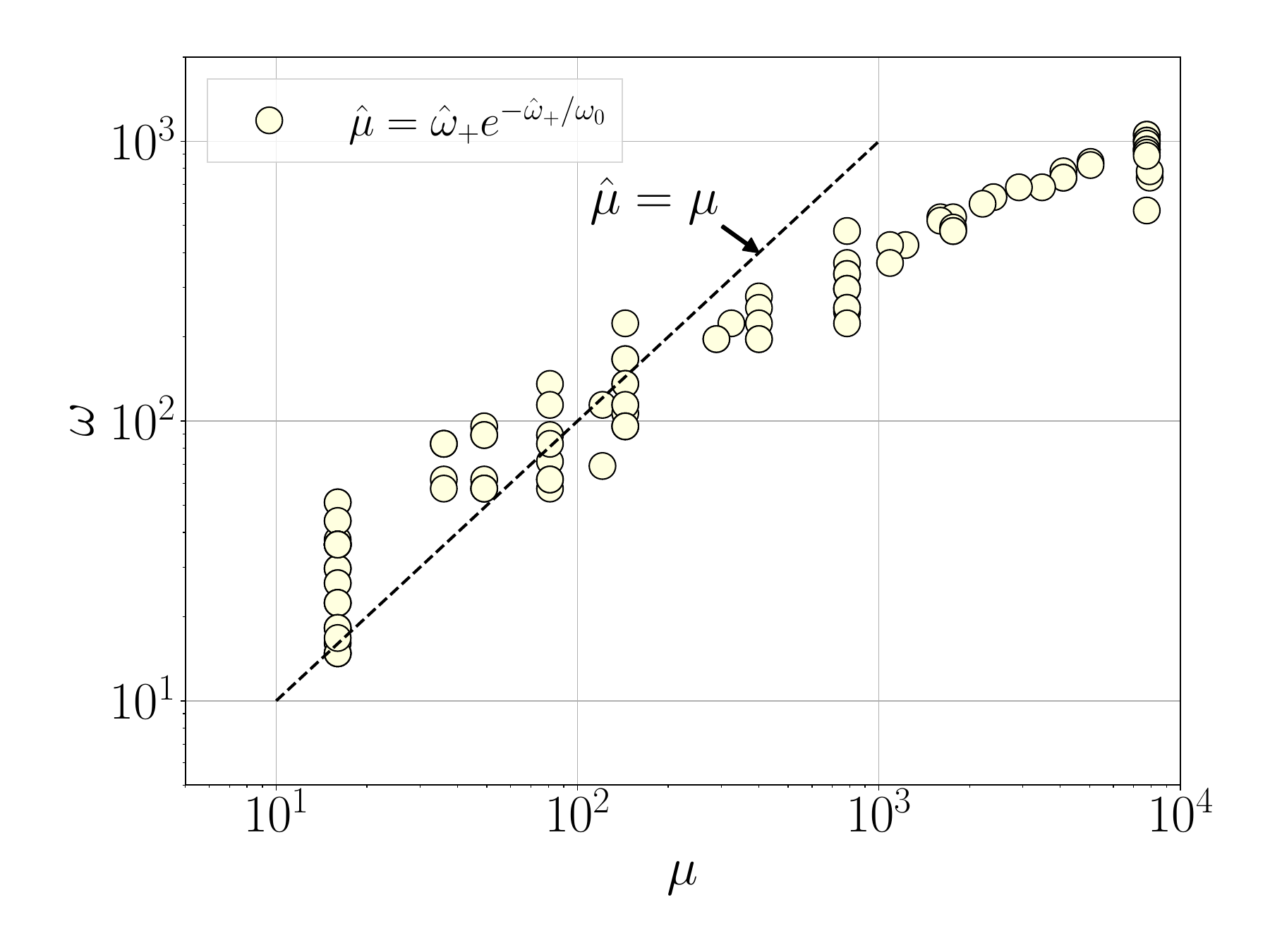}
\caption{(Left: DAM and Right: NLS). Scaling of the Rayleigh-Jeans  potential estimate  $\hat \mu$ and left fronts $\omega_-, \hat{\omega}_-$ (for DAM only) with respect to the estimate $\mu$.}
\label{fig:3}
\end{figure}
Figure~\ref{fig:2} shows the dynamical evolution of the estimates~\eqref{eq:muT} and~\eqref{eq:muT-cons}, together with the front estimate $\hat{\omega}_+$ for both the NLS and the DAM. The frequency $\hat{\omega}_+$ is defined such  that the energy spectrum drops at a certain threshold fraction $\sigma$ of its maximum value, \emph{e.g.},  
$E(\hat{\omega}_+)= \sigma \max_{\omega}  E_\omega$, with $\sigma=0.4$ for the DAM and $\sigma=0.7$ for the NLS.  The thresholds are selected so that both estimates~\eqref{eq:muT} and~\eqref{eq:muT-cons} are as close as possible. For both the DAM and NLS, we observe the algebraic propagation $\hat{\omega}_+(t) =C_+ t^{1/3}$ for the RJ front, where $C_+$ is a constant $\simeq 5\omega_0$  for DAM and  $\simeq 6\omega_0$ for NLS. For DAM, the right-front also grows algebraically $\omega_+(t) \propto t^{1/3}$ but with a different prefactor and as we will see later, this growth is predicted by self-similarity arising from energy conservation. 
Combined with Eq.~\eqref{eq:muT-cons}, this leads to  the cooling laws
\begin{align}\label{eq:coolinglaw}
	\hat T = \dfrac{\mathcal E}{C_+}t^{-1/3}\quad   \ln \left(\hat{\mu}\right)\sim -\dfrac{C_+}{\omega_0}t^{1/3}; 
\end{align}
which are  observed  both for NLS  and the DAM. Fig.~\ref{fig:2} and \ref{fig:3} show that the temperature and potential estimates are consistent with each other, e.g., $ T(t)\approx\hat T(t), \mu(t) \approx  \hat{\mu}(t)$ both for the DAM and for NLS.
Note that the law for the logarithmic potential stems from the  more general estimate $ \hat{\mu} \sim C_+t^{1/3}  e^{-C_+t^{1/3}/\omega_0}$, which is also verified in Fig.~\ref{fig:2}.

For the DAM, the left fronts have nontrivial dynamics.  We estimate the RJ fronts by thresholding the wave action to a fraction $\tilde \sigma=  0.4$ of its maximum. The left panel of Fig.~\ref{fig:3} shows  $\omega_-$ and  $\hat \omega_-$ versus $\mu$, revealing  scaling  relations $\hat \omega_- \propto  \mu^{2}$  with exponent $\beta\simeq 2$. This  implies  $\ln \left(\hat \omega_-\right) \propto -t^{1/3}$, similar to the cooling law for the logarithm of the potential,
The behaviors of $\mu(t), \omega_-(t), \hat \omega_-(t)$ are also related to self-similarity, although of a different kind than that governing for  right front---we postpone the discussion until \S \ref{sec:SSDAM}. 
For NLS,  recall that no sharp left front  exists due to the nonlocality of the interaction described by the collision term of the WKE.
\begin{figure}
\includegraphics[width=\textwidth]{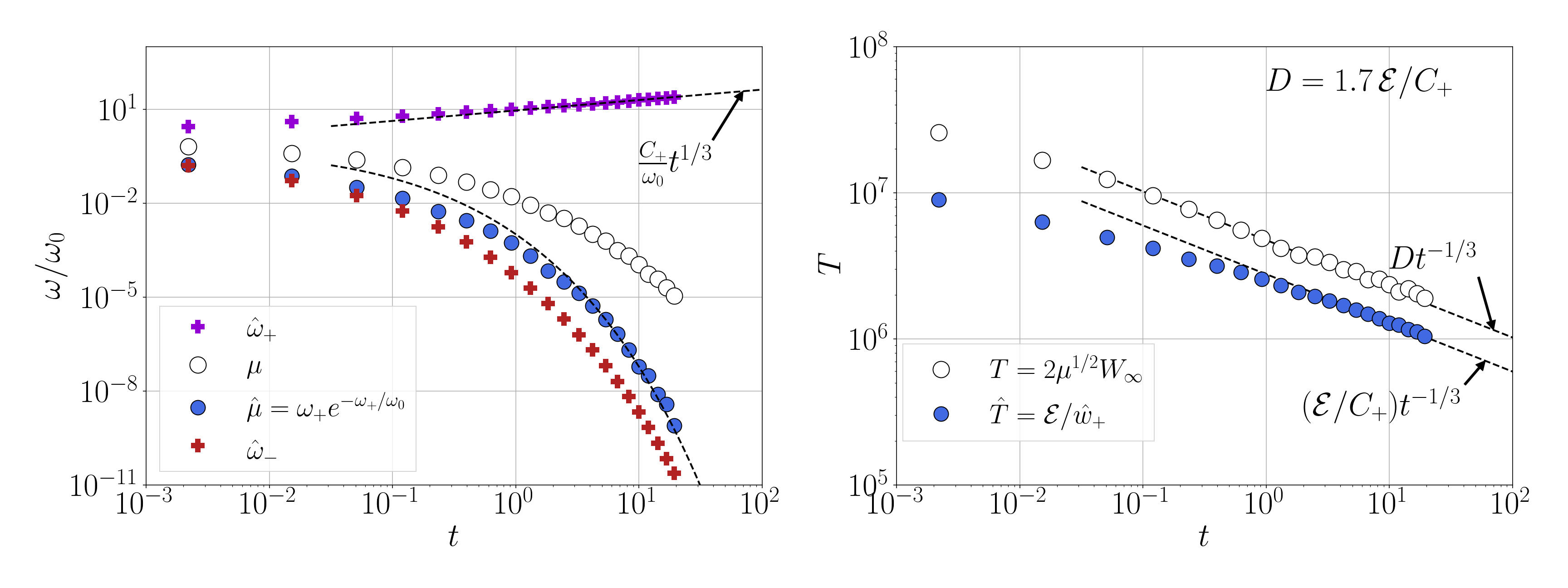}\\
\includegraphics[width=\textwidth]{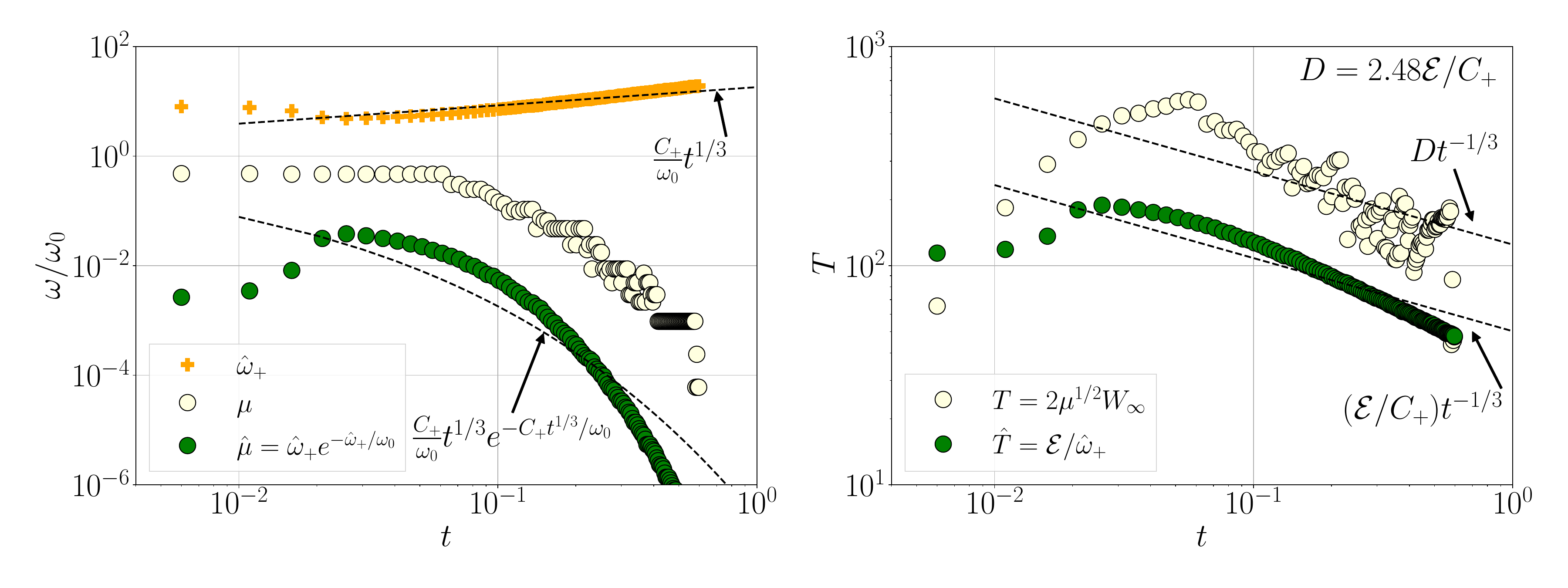}
\caption{
(Top: DAM and Bottom: NLS). Same as Fig.~\ref{fig:2} but now with the right front frequency $\hat{\omega}_+$ estimated using $0.2$ of the maximum energy spectrum for both the DAM and the NLS.}
\label{fig:2a}
\end{figure}

By changing the threshold values delimiting the RJ range, discrepancies arise between various temperature estimates.
For example, Figure~\ref{fig:2a} shows the value $\sigma =0.2$ for both DAM and NLS. For DAM, we now see a factor of two discrepancies between the temperature measurements, although the temporal scaling remains identical $T(t)  \approx 2 \hat{T}(t) \propto t^{-1/3}$. The two chemical potential measurements are no longer proportional to each other, but they still obey a quite robust scaling $\mu (t) \propto \hat{\mu}^{b'} (t)$, $b' \simeq 0.5$ (not shown). This means that the log-potentials share the same temporal scaling $\ln \mu(t) \approx 0.5 \ln \hat{\mu}(t)  \propto -t^{1/3}$, up to a non-universal prefactor. For the NLS, we observe significantly larger discrepancies between the estimates~\eqref{eq:muT} and~\eqref{eq:muT-cons}, reaching two orders of magnitude, both for the temperature and the chemical potential. However, and up to this prefactor, the algebraic rate $\propto t^{-1/3}$ is essentially unchanged. Note that upon further reducing the threshold to $\sigma \lesssim 0.1$, the temporal scaling $t^{1/3}$ disappears significantly in the NLS. We attribute this feature to the presence of a dissipative mechanism combined with nonlinear interactions.

The prefactor discrepancy highlights the fact that, strictly speaking, the high-frequency part of the profile is not 
prescribed by RJ during relaxation. Our estimation of $\hat T$ and $\hat \mu$ are based on the assumption that the shape of the spectrum is RJ up to a sharp front $\hat{\omega}_+$ and  are  by construction sensitive to the spectrum threshold that defines the position of $\hat{\omega}_+$.  This non-universality is captured by the variations of the prefactor $C_+$. For the DAM, the latter increases from $\approx  5 \omega_0$ at $\sigma =0.4$  up to $ \approx 30$ at $\sigma = 10^{-15}$.

\section{Self-similarity of the high-frequency spectrum}
\label{sec:SSHF}
\noindent
\noindent
At high-frequencies, discrepancies with RJ occur as  $\hat{\omega}_+ <\omega <\omega_+$, where $\omega_+$ is the position of the right front of the spectrum.   Both for DAM and NLS ,  this discrepancy comes from the emergence of a self-similar front.
To evidence this self-similarity, this section  recalls  a classical calculation.
We assume that the self-similar solution  matches the high-frequency asymptotics of the RJ spectrum, $N_\omega = T/\omega$, on its left (low-frequency) side and  seek a similarity spectrum  in the form 
\begin{align}
N_\omega = A\, t^{a} \, f(\eta) \quad
\text{with} \quad \eta = \frac{\omega}{\omega_+(t)}, \quad
\omega_+(t) = C_+ t^b, \quad C_+ =\text{const}>0.
\end{align}
Substituting this expression into the WKE and DAM and requiring that the resulting equations for $f(\eta)$ involve only the similarity variable $\eta$, and not $t$ or $\omega$, we get the condition $b=-2a -1$ and
\begin{align}
\label{eq:ssDAM}
   a  f +(2a+1) \eta  \frac{d f}{d \eta} &=  A^2 C_+ \frac{d^2}{d \eta^2}\left( \eta^5 f^4 \frac{d^2}{d \eta^2}\left( \frac{1}{f}\right)\right) \quad& \text{for DAM,} \\
   a  f +(2a+1) \eta  \frac{d f}{d \eta} &= A^2 C_+ \, 
   St[f] \qquad &
 \text{for WKE,}
 \label{eq:ssWKE}
\end{align}
where the collision integral $St[f]$ is given by the RHS of~\eqref{eq:kinetic_omega} in which $\omega_i$ and  $N_{\omega_i}$ are replaced by $\eta_i$ and $f({\eta_i})$.

To formulate the problem of finding the self-similar profile $f(\eta)$, one must complete it with boundary conditions on both sides.
As mentioned above, the self-similar solution must match the RJ tail $N_\omega = T/\omega$ on the left side. This gives the boundary condition $f(\eta) \sim 1/\eta$, as $\eta \to 0$ and the identification 
\begin{equation}
\label{eq:temperatureiden}
T(t)=AC_+t^{a+b} .
\end{equation}
On the right side, a natural condition for both  DAM and WKE is that the profile has a rapidly decaying end $f(\eta) \to 0$ at $\eta \to \infty$ fast enough for $P$ and $Q$ defined in~\eqref{eq:nDAM} and~\eqref{eq:eDAM}, to $P, Q \to 0 $ as $\omega \to \infty$. In other words, the right portion of the spectrum is ``particle-and-energy proof".

Energy conservation drives the right front dynamics and prescribes
${\cal E} = \int_{\mathbb{R}_+} \omega N_\omega \, \de\omega = A C_+^2 t^{a+2b}  \int_{\mathbb{R}_+} \eta f(\eta) \, d \eta =  \hbox{const} $. This  gives  the additional condition $a=-2b$ and subsequently
\begin{align*}
a=-2/3, \;\; b=1/3, \quad \longrightarrow \quad \omega_+(t) = C_+ \, t^{1/3},
\end{align*}
which is exactly the right front scaling observed numerically as reported in the previous section.
The RJ tail $N_\omega = T/\omega$ on the left side makes the waveaction non-integrable. 
From the identification \eqref{eq:temperatureiden}, it however 
implies that $T \propto t^{-1/3}$, hence recovering  the cooling law \eqref{eq:coolinglaw}.

\section{Self-similar solutions in the DAM}
\label{sec:SSDAM}

\noindent We will now investigate the partial thermalization and cooling dynamics from the perspective of blowups associated with the self-similar solutions. In this section, we will restrict our analysis to the DAM, as it allows access to a much wider range of scales and avoids noise that would otherwise be in the 2D NLS simulations. In the following section, we will discuss parallel results for the NLS. 

The concept of self-similarity exists in several forms that are connected to the blowup behaviors of the physical system~\cite{barenblatt1996scaling}. A standard dichotomy distinguishes between, on the one hand, first-kind self-similarity in which the spectrum evolves into the UV range over an infinite time at a rate determined by conservation laws, and second-kind self-similarity, on the other hand, associated to finite time blow-up, and anomalous scaling laws. There is also a third kind of self-similarity discovered in~\cite{Bell_2018,Nazarenko_2019} that does not fit into the standard classification. There, the similarity scalings are inherited from an adjacent self-similar region rather than from the conservation laws (like in the first-kind) or determined by solving a nonlinear eigenvalue problem (like in the second-kind). We will see that in our case the various portions of the spectrum evolve following different kinds of self-similar solutions, characterized by different blowup rates.

\subsection{General framework}
\label{sec:framework}
Let us first  outline  a general scheme that connects self-similar solutions to blowup classifications--whether the latter occur in finite time or not. The first step consists in identifying a norm that diverges over a finite (or infinite) time window, say at a rate $\mW_g(t)\to\infty$. The idea is to reparametrize time in terms of this blowup rate, in order to map self-similar solutions of Eq.~\eqref{eq:dam} onto asymptotic traveling-wave solutions of a suitably rescaled autonomous dynamics~\cite{giga1987characterizing,gilson1998two,eggers2008role,mailybaev2012renorm,schubring2026blowup}.
In our case, the rescaling schemes will be determined by the asymptotic (diverging) behavior of the norms
\begin{align}\label{eq:norm}
	|W_g|_\infty \sim \mW_{g}(t),\quad W_g (\omega,t) := \omega^{g+1/2} N_\omega.
\end{align}
Depending on its sign, the parameter $g \in \mathbb R$ either weights the UV ($g>1/2$), the central  $(-1/2<g<1/2)$ or the infrared (IR) ($g<-1/2$) frequencies. For $g=0$, the profile $W_g(\omega,t)$ recovers the symmetric profile shown in Fig.~\ref{fig:1} and previously introduced Eq.~\eqref{eq:muT}. By analogy with UV blowup criteria in classical hydrodynamics~\cite{beale1984remarks,bustamante2012interplay}, one can think of $W_g(\omega,t)$ as a generalized vorticity, whose blowup rate determines the function $\mW_g(t)$. 
From \S\ref{sec:kinematics}, we know that the profiles are closely related to RJ spectra in a finite but increasing frequency range  $\hat{\omega}_- (t) < \omega < \hat{\omega}_+ (t)$. We therefore expect three different phases, depending on the norm~\eqref{eq:norm} diverging because of IR scales, bulk or UV scales. This trichotomy leads to various forms of self-similar solutions, describing self-similar dynamics of the left and right fronts, as well as that of the bulk portion of the profile.

To be more explicit, let us observe that the generalized vorticities obey the dynamics
\begin{align}\label{eq:gdynamics}
	\partial_t W_g(\omega,t) =\omega^{-2g} F_g[W_g],\quad F_g[W]:=D_{1/2-3g}D_{3/2-3g}W^4 D_{g-1/2}D_{g+1/2} W^{-1},
\end{align}
where $F_g$ is a homogeneous functional of degree three: $F_g(\lambda W) = \lambda^3 F_g(\lambda W)$, built from the differential operators
\begin{align*}
	D_\alpha W:= \alpha W +  \omega \partial_\omega W.
\end{align*}
Our strategy is to rescale the vorticity dynamics~\eqref{eq:gdynamics} by introducing the variables
\begin{align}\label{eq:rescaledOm}
	\Omega (\kappa,\tau) = \dfrac{W_g}{ \mW_g(t)} \,\quad \tau =  \ln\left(\mW_g\right)  \quad \kappa= \ln\left(\omega\right).
\end{align}
This leads to the rescaled vorticity dynamics~\eqref{eq:gdynamics}
\begin{align}\label{eq:omega}
	\partial_\tau \Omega(\kappa,\tau) =-\Omega + \dfrac{\mW_g^3}{\dot \mW_g} e^{-2g\kappa }F_g[\Omega],
\end{align}
where the dot denotes differentiation with respect to the (unrescaled) time $t$. Self-similar solutions refer to traveling wave solutions of Eq.~\eqref{eq:omega}; their existence and properties depend on the dynamical behavior of $\mW_g$, and more specifically on whether the prefactor
\begin{align}\label{eq:phi}
	\dfrac{\mW_g^3}{\dot \mW_g} e^{-2g\kappa},
\end{align}
can be expressed as a function of the variable $\kappa -c\tau$ for $c$ a yet-to-be-specified traveling velocity in log-frequency space.

\subsection{UV blowup and right-front self-similarity}
\begin{figure}
\includegraphics[width=0.49\textwidth]{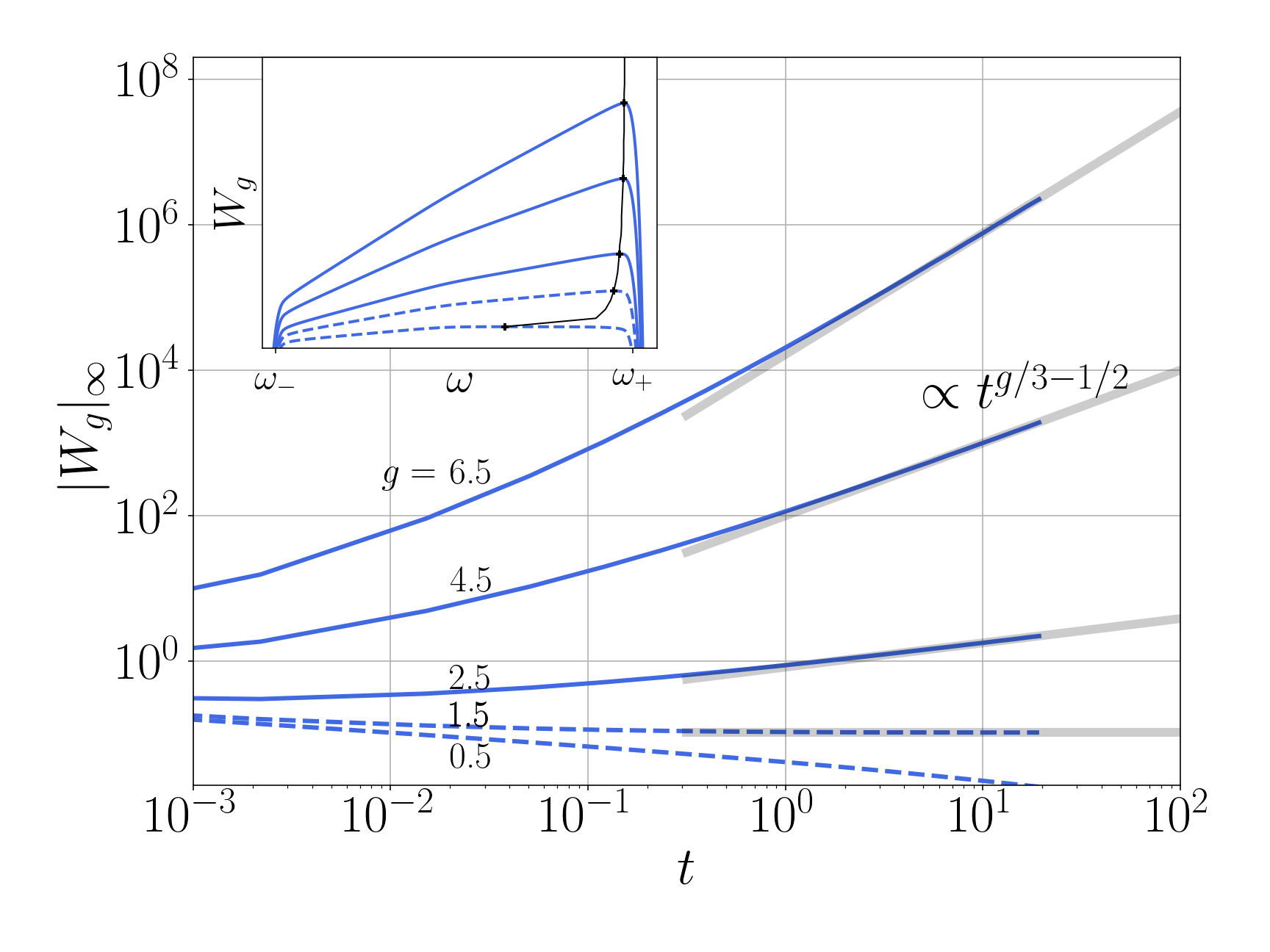}
\includegraphics[width=0.49\textwidth]{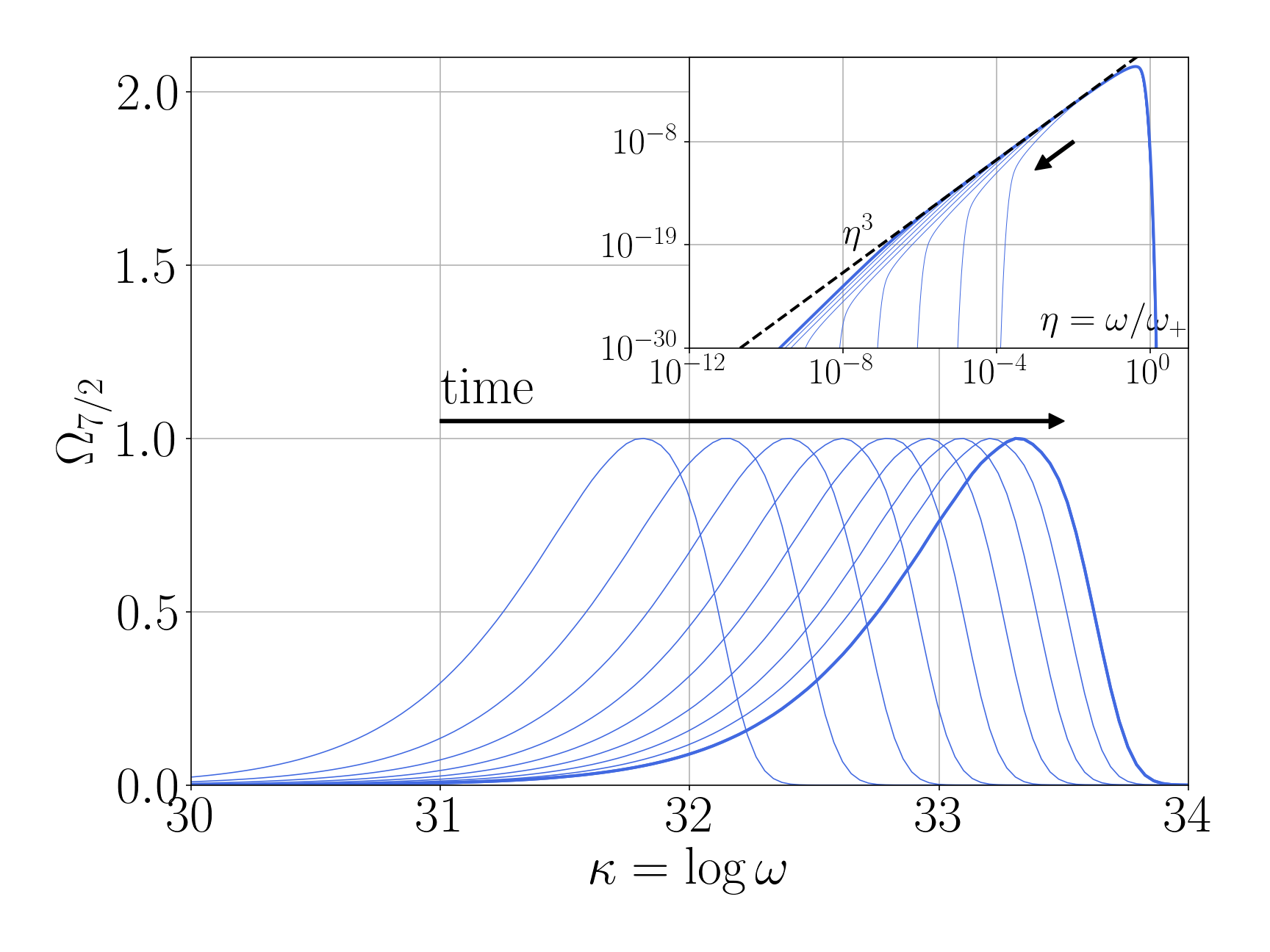}
\caption{UV blowup. Left panel shows the temporal evolution of  $|W_g|_\infty$ for $g \ge 0$, normalized by their initial values. Solid lines indicate  $g \ge 3/2$. Inset shows the corresponding profile of $W_g$ with the black line tracking the maximal values. Right panel: traveling wave solution  of Eq.~\eqref{eq:omega} observed for $g=7/2$. Inset shows the corresponding self-similar profile in log-log coordinates.\label{fig:4}}
\end{figure}

\noindent Let us first re-consider the right-front (UV) dynamics in terms of $W_g$. It connects to the algebraic divergence of the norms $|W_g|_\infty$ for any $g>1/2$, occurring at a rate  $\propto t^{g/3-1/2}$--see Fig.~\ref{fig:4}. The interpretation is straightforward: for the UV end to drive the blowup, the weight $\omega^g$ in Eq.~\eqref{eq:norm} must sufficiently amplify the UV range--see the inset in the left panels. 
As the algebraic behavior $\propto t^{g/3-1/2}$ follows directly from Eq.~\eqref{eq:rescaledOm}, together with energy conservation, which is essentially the approach we followed in \S\ref{sec:SSHF} dealing directly with the spectrum $N_\omega$. Then, the right-front UV dynamics corresponds to first-kind self-similarity. The calculation proceeds as follows. On the one hand, assuming a power law behavior $ \mW_g =A_g  t^{\alpha}$ in Eq.~\eqref{eq:phi} yields
\begin{align*} 
	\dfrac{\mW_g^3}{\dot \mW_g} e^{-2g\kappa }  = \phi(\kappa-c\tau),\quad \phi(y) :=    \dfrac{A_g^2}{\alpha} e^{-2g y},\quad c:= \dfrac{1}{g}\left(1+\dfrac{1}{2\alpha}\right).
\end{align*}
This suggests that Eq.~\eqref{eq:omega} can support, at least asymptotically, traveling wave solutions of the form  $\Omega(\kappa,\tau) = \Psi(\kappa-c\tau)$. In the original variables, this translates into  convergence towards self-similar  profiles $\psi$ as
\begin{align}\label{eq:ssWg}
	W_g(\omega,t) \sim \mW_g(t) \Psi\left(\eta\right),\quad \eta=\frac{\omega}{\mW^c}, \quad \psi(\eta) = \Psi\left[\ln\left(\eta\right)\right].
\end{align}
On the other hand, the energy and wave action formally asymptote to
\begin{align}\label{eq:EN-ss}
	\mN \sim \mW_g^{1+c(1/2-g)}  \int_{\mathbb R^+}  \eta^{-g-1/2} \psi\left(\eta\right)\ \de\eta,\quad \mE \sim \mW_g^{1+c(3/2-g)}  \int_{\mathbb R^+}  \eta^{1/2-g} \psi\left(\eta\right)\ \de\eta;
\end{align}
However, as we explain in \S\ref{sec:SSHF} the conservation of wave action has to break down. Indeed, from Fig.~\ref{fig:1}, we observe that the self-similar profile connects to the RJ asymptotics $N(\omega,t) \propto 1/\omega$ observed in the intermediate range  $\mu \ll \omega \ll \omega_+$. This suggests that the profile satisfies $\eta_{-g-1/2} \psi(\eta) \simeq \eta^{-1}$ for $\eta \ll 1$, implying a divergence at $0$ in the integral involved in $\mN$, but convergence for the one involved in $\mE$. In other words, wave action conservation breaks down, whereas energy conservation does hold, and prescribes 
\begin{align}\label{eq:ccons}
	c = \dfrac{1}{g-3/2}. 
\end{align}
Combined with Eq.~\eqref{eq:ssWg}, this yields $\alpha =g/3-1/2$,  and the late-time asymptotics $\mW_g(t) = A_gt^{g/3-1/2}$. As the weighting parameter $g$ increases, the vorticity profile becomes increasingly peaked at the right front $\omega_+$. This leads to the scaling estimates
\begin{align*}
\omega_+ \sim \mW_g^{1/g}(t) \propto t^{1/3},\quad \text{for}\quad g\gg1,\quad t\gg 1,
\end{align*}
consistent with the scaling reported in Fig.~\ref{fig:2}. The convergence towards the traveling wave solution and associated self-similar profile is illustrated in the right panel of Fig.~\ref{fig:4} using the parameter $g=7/2$.

\subsection{IR blowup and left-front self-similarity}
\begin{figure}
\includegraphics[width=0.49\textwidth]{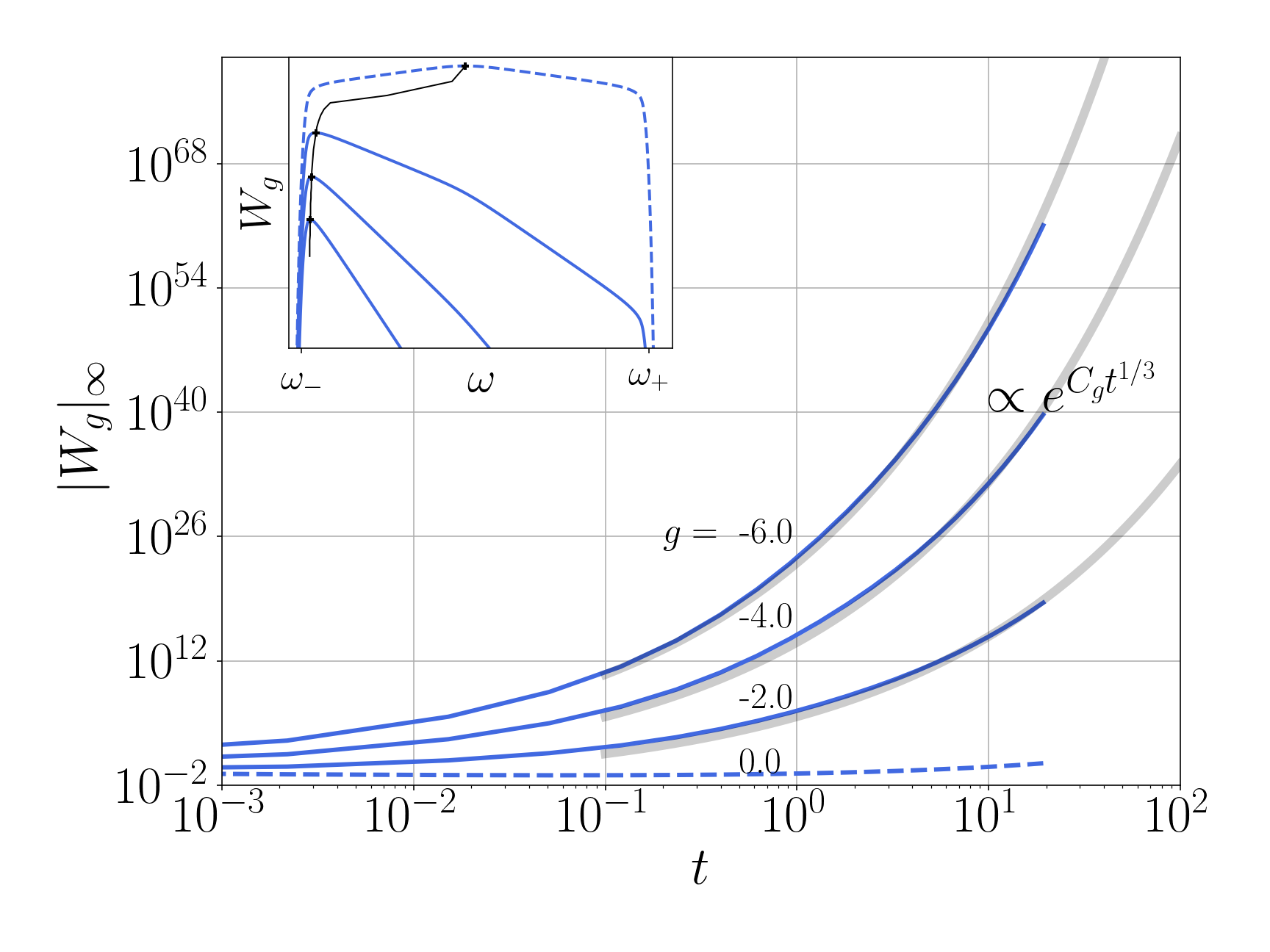}
\includegraphics[width=0.49\textwidth]{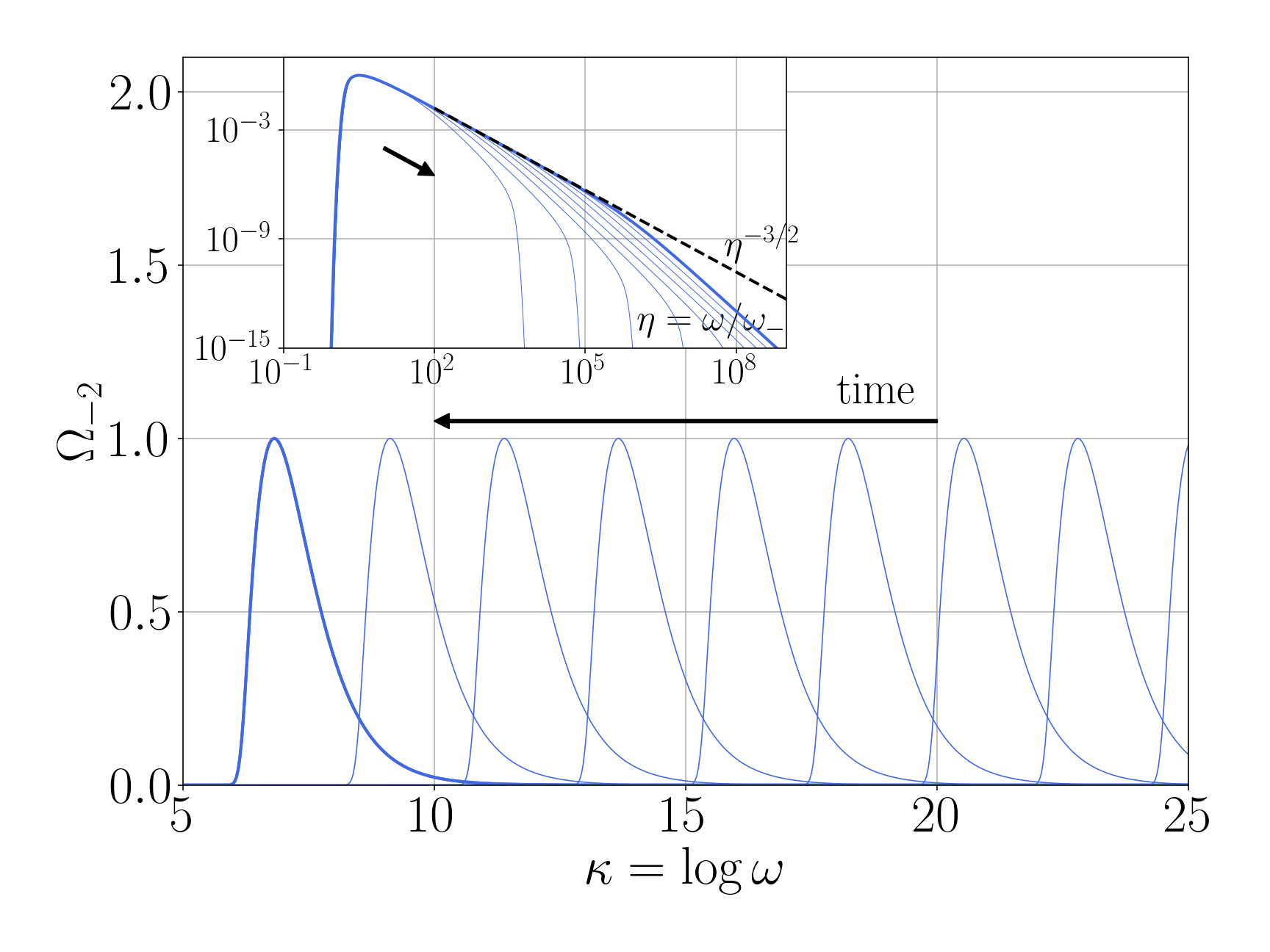}
\caption{IR blowup. Same as in Fig.\ref{fig:4}, but for negative $g$. Solid lines in the left panels indicate  $g \le -1/2$ and right panel uses $g=-2$.}
\label{fig:5}
\end{figure}
Similarly, the left front dynamics drives the divergence of the norms $|W_g|_\infty$ for $g\le -1/2$  obtained by  weighting the IR frequencies. From Fig.~\ref{fig:5}, we characterize this divergence as a stretched exponential:
\begin{align}\label{eq:wgminus}
	|W_g|_\infty \propto  e^{C_g t^{1/3}} , \quad C_g \simeq 9 |g|,\quad \text{for  $g\le 1/2$}.
\end{align}
Here, the symbol $\propto$ means up to a  multiplicative factor--possibly non constant--that remains dominated by the stretched exponential term. The divergence of Eq.~\ref{eq:wgminus} dictates the left front behavior: For strongly negative $g$ the norms $|W_g|_\infty$ are dominated by $\omega_-$, such that 
\begin{align*}
	 \ln\left(|W_g|^{1/|g|}_\infty\right)  \sim |\ln\left(\omega_-\right) | \propto t^{1/3}, \quad \text{for $g\ll -1$}.
\end{align*}

The inset of Fig.~\ref{fig:5} shows evidence that the IR end converges towards a self-similar profile essentially characterized by a sharp IR cutoff connecting to the constant portion of the RJ solution $N(\omega, t) \propto 1$, observed in the range  $\omega_- \ll\omega \ll \mu$. In particular, this sets the asymptotics $\psi(\eta) \simeq \eta^{g/2}$ for $\eta =\omega/\omega_-(t)\gg 1$. Similarly to the sharp right front, fast decay of the sharp left front ensures that the left boundary is (like the right boundary) also ``particle- and energy-proof": $P, Q \to 0$ for $\omega \ll \omega_-$


As such, the integrals in Eq.~\eqref{eq:EN-ss} diverge towards the UV and the blowup features are not constrained, neither by the conservation of wave action nor energy--unlike for the UV front. This makes the IR self-similarity different from the first-kind. Neither is it of the form of a second-kind self-similarity as it develops over an infinite time. We will argue later that the IR self-similarity inherits its scaling from matching to another (central part) self-similar solution--a property named third-kind self-similarity in Ref.~\cite{Bell_2018,Nazarenko_2019}.

To explicitly connect this self-similar profile to a traveling wave solution of Eq.~\eqref{eq:omega}, one needs to discard spurious multiplicative factors. Explicitly, from the prescription  $\mathcal{W}_g(t) \propto  e^{C_g t^{1/3}}$, we estimate  $\dot \mW_g \propto \mW_g(t)$, and consequently 
\begin{align}\label{eq:irestimate}
\dfrac{\mW_g^3}{\dot \mW_g} e^{-2g\kappa } = \phi(\kappa-c\tau),\quad \phi(y) \propto   e^{-2g y},\quad c:= \dfrac{1}{g}.
\end{align}
As $g<0$, the travel speed $c$ is negative, which means that the front propagates to the IR end, as it should be. We emphasize that the crucial point of the estimate~\eqref{eq:irestimate} is to disregard any prefactor in $\phi$ that remains dominated by the stretched exponential. While, in principle, we  expect the traveling wave to appear only in the asymptotics $t\gg 1$,  the right panel of Fig.~\ref{fig:5} shows that it can be clearly identified as soon as the constant portion of the RJ spectrum extends about two decades or so.

\subsection{Central blowup and deviations from RJ}
\begin{figure}
\includegraphics[width=0.49\textwidth]{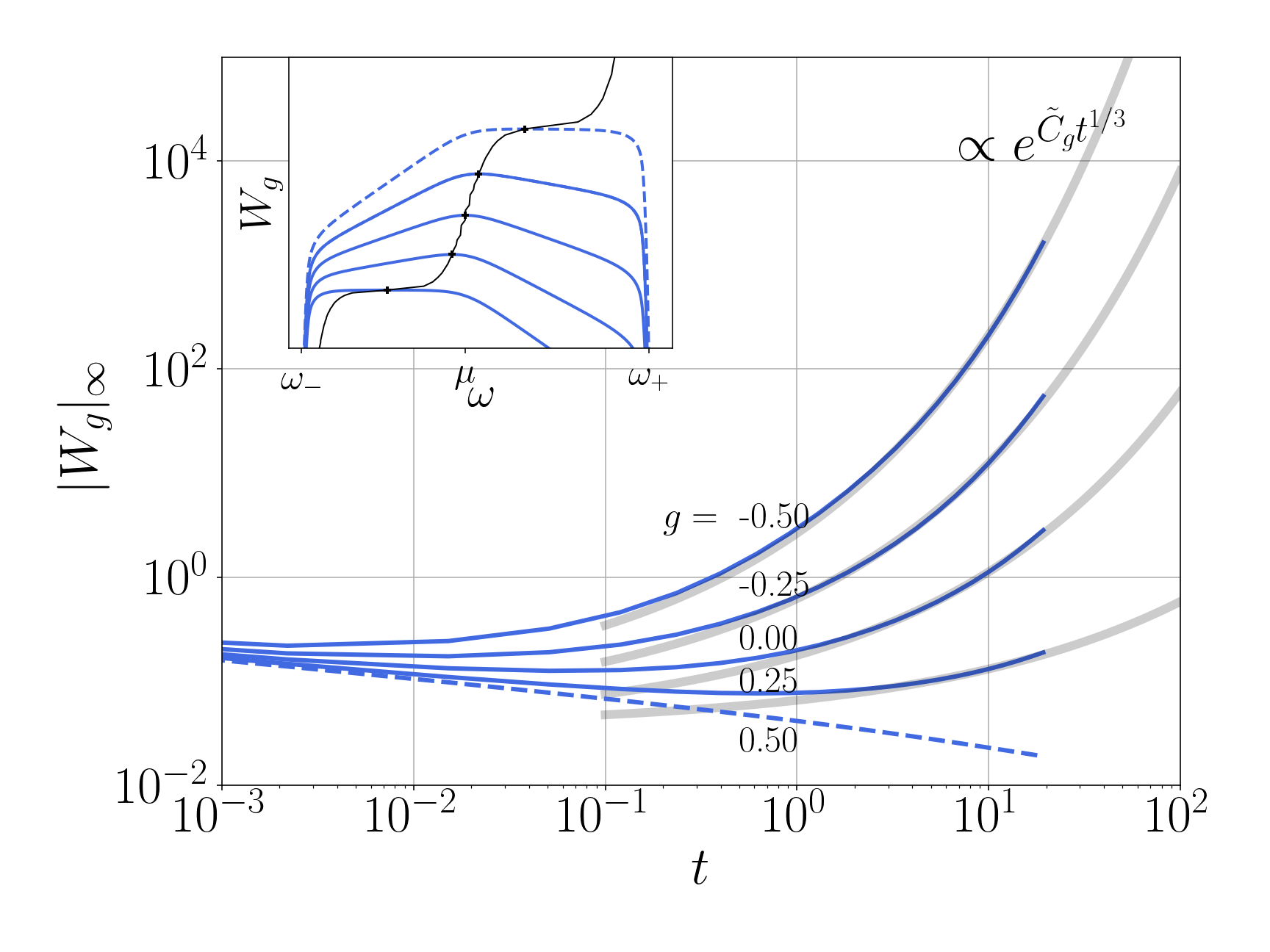}
\includegraphics[width=0.49\textwidth]{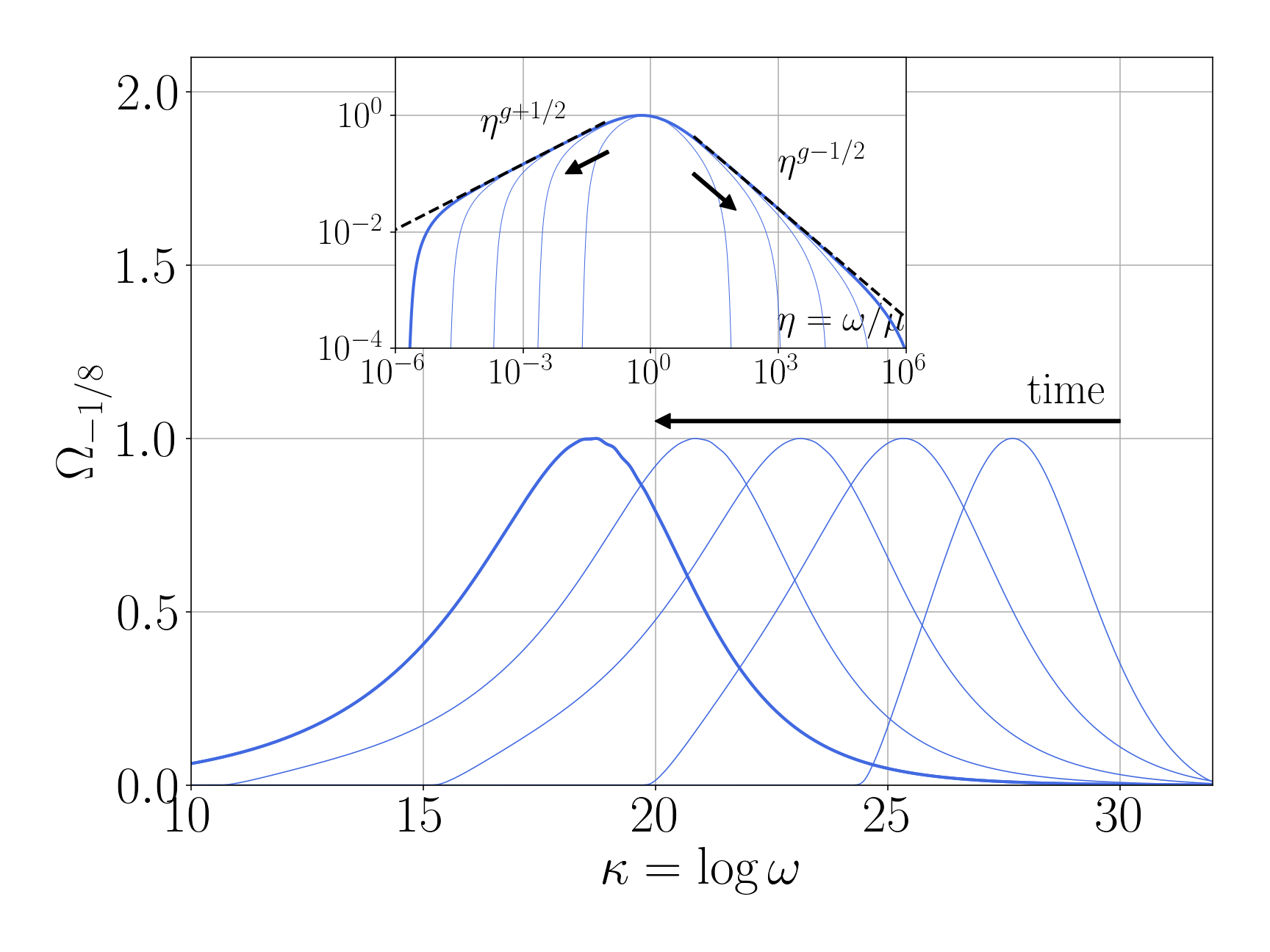}
\caption{Bulk blowup. Same as in Fig.\ref{fig:4}, but for  $g \in (-1/2,1/2)$. Solid lines in the left panels indicate $g \le 1/4$ and right panel uses $g=-1/8$, corresponding to a slight tile of the symmetric profile in the inset of Fig.\ref{fig:1}. 
\si}
\label{fig:6}
\end{figure}
We can now examine the central region, which governs the norms $|W_g|_\infty$ for $g \in (-1/2,1/2)$, implying a slight tilt of the symmetric profile $W=\omega^{1/2}N_\omega$.  

The dynamics of the central region turns out to be analogous to the IR dynamics. Fig.~\ref{fig:6} shows that for any $g \in (-1/2,0)$, the norms  diverge with a stretched-exponential rate $|W_g|_\infty \propto  e^{\tilde C_g t^{1/3}}$. This is akin to the IR case, except that the prefactors $\tilde C_g$ differ from the asymptotic linear scaling in $g$ reported in Eq.~\eqref{eq:wgminus}, i.e., $\tilde C_g  >  9|g|$. The divergence of those moderate-$g$ norms reflects the dynamics of the potential $\mu$ measured in Fig.~\ref{fig:2} as the argmax of the symmetric profile $W(\omega,t)= \omega^{1/2}N_\omega$--see Fig.~\ref{fig:2} and Eq.~\eqref{eq:muT}. As such, the norms reflect the dynamics of the ``bulk" region of the RJ distribution. For $g<0$, we may therefore repeat the calculation of the IR case and conclude that Eq.~\ref{eq:irestimate} holds, prescribing a solitonic propagation of the bulk towards the IR at speed $1/g<0$.

The right panel of Fig.~\ref{fig:6} shows the case $g=-1/8$ and displays that both the left and right tails follow RJ asymptotics, yielding the natural question: Is the self-similar profile precisely RJ? The answer must be negative. To see this, note a direct contradiction. Recall that the rescaled profile $\Omega(\kappa,\tau)= W_g/|W_g|_\infty$  evolves with Eq.~\eqref{eq:omega} as 
%
$\partial_\tau \Omega(\kappa,\tau) =-\Omega + \left(\mW_g^3/\dot \mW_g\right)e^{-2g \kappa  } F_g[\Omega].$
On the one hand, a traveling $RJ$ solution, e.g., $\Omega(\kappa,\tau)= \Psi_{RJ}(\kappa-\tau/g)$ would be prescribed by the explicit formula (valid for $|g|<1/2$):
\begin{align*}
	\Psi_{RJ}(\sigma) = Z_g e\dfrac{e^{\sigma(g+1/2)}}{e^\sigma+1}, \quad Z_g = \dfrac{g+3/2}{(1/2-g)^{g+1/2}},
\end{align*}
where $Z_g$ normalizes the solution so that its maximum is 1. On the other hand, it prescribes $F_g[\Psi_{RJ}] = 0$ in Eq.~\eqref{eq:omega} , resulting in  $\Psi'_{RJ} -g\Psi_{RJ} =0$. This prescribes a pure scaling solution $\Psi_{RJ}(\sigma)= e^{g\sigma} =\eta^g$, obviously different from the subsumed RJ profile.

Actually, there is an even simpler argument as to why the observed self-similar profile cannot be exactly RJ in any finite frequency window. Indeed, RJ is an exact stationary solution, which means that the values of the temperature and the chemical potential should not evolve in time--contrary to what is observed in our system. 

Let us conclude this analysis by a qualitative comment related to  the  cooling law \eqref{eq:coolinglaw}  discussed in \S\ref{sec:kinematics}.
The central part of the profile is close enough to the RJ shape to be able to characterize it in terms of the varying temperature and chemical potential. 
To obtain the cooling law $\ln\omega_-(t) \propto -t^{1/3}$, we need to make an additional assumption that the central and the left ranges are two parts of the same self-similar solution. Indeed, the similarity variable of the left range is $\eta=\omega/\omega_-(t)$ and, since the RJ denominator is $\omega +\mu$, it can be part of the same self-similar solution only if $\omega/\mu \propto \eta=\omega/\omega_-(t)$, i.e. $\omega_-(t) \propto \mu(t)$ and $\ln\left(\omega_-(t)\right) \propto \ln\left(\mu(t)\right) \propto - t^{1/3}$, as required. Of course, this argument is only qualitative since the left and center ranges cannot be part of the same self-similar solution exactly, as seen, e.g., from the difference between $C_g$ and $\tilde C_g$. On the other hand, this argument allows us to understand that the similarity scaling of the left solution arises from matching it to the central solution, a behavior discovered in Ref.~\cite{Bell_2018,Nazarenko_2019}. (In fact, Ref.~\cite{Nazarenko_2019} considered a self-similar formation of the thermodynamic state by a front propagating toward the IR end of the spectrum in the Leith model of hydrodynamic turbulence---a problem very similar to the left-propagating front considered in the present paper.)


\section{Self-similar solutions in NLS}
\label{sec:SSNLS}

The freely evolving NLS can be interpreted in a manner similar to DAM. However, there is an important difference from the DAM: the NLS evolution does not show any left front: the spectrum fills in the entire low-$k$ region up to $k=0$ almost instantaneously. This can be understood by analyzing the  evolution described by the WKE dominated by non-local interactions, which predicts that the self-similar profile must tend to a constant function at small $\eta $. This property of our 2D solution is shared by the 3D self-similar solutions previously considered in~\cite{SEMISALOV2021105903,zhu2023self}. However, this also means that in the WKE, the IR and the central (RJ) parts are naturally combined into the same self-similar solution, which (like in DAM) appears to be of the third-kind. Numerical simulations show evidence of first-kind self-similarity for the right front, and the third-kind self-similarity for the bulk (including the IR part).  This section  briefly comments on these regimes in the NLS evolution examined in terms of the same blowup norms as we used for the DAM.

\subsection{UV  front}
\begin{figure}
\includegraphics[width=\textwidth]{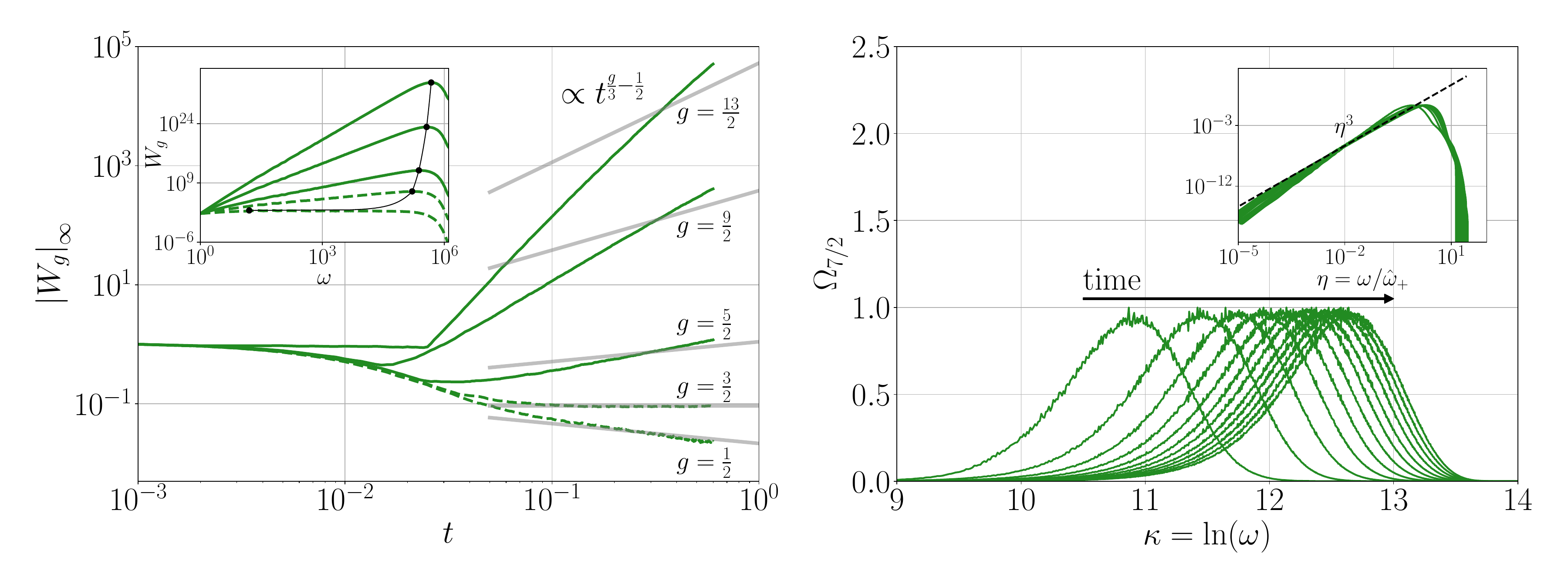}
\caption{Same as in Fig.\ref{fig:4} but for NLS.}
\label{fig:7}
\end{figure}
Compared to Fig.~\ref{fig:4} for the DAM, Fig.~\ref{fig:7} presents numerical evidence that in 2D NLS, the suprema of the profile $W_g$, although algebraically divergent, does not follow the law $\propto t^{g/3-1/2}$ for $g>3/2$. On the other hand, $W_{3/2} \propto const$ which is consistent with the self-similarity of the first kind and the results presented in Fig.~\ref{fig:2}. This is evidence of the fact that the high-frequency part of the UV tail, whose contribution is magnified for large $g$, does not follow the predicted scaling, most likely due to the finite resolution of the NLS simulations, effects of the hyper-viscosity and the finite wavenumber range. This conclusion is consistent with the deviations from the predicted self-similarity when the right front is chosen at a low threshold, as seen in Fig.~\ref{fig:2a}.

The right panel of Fig.~\ref{fig:7} specifies the case $g=7/2$ and displays the convergence of $W_{7/2}$ towards a solitonic wave in the log-frequency space $\kappa =\ln\left(\omega\right)$. This solution is prescribed by a universal profile $\Omega_{7/2}$ characterized by a sharp UV front and asymptotic behavior $\propto \eta^{g-1/2} \propto \eta^{3}$, indicating equipartition of energy in the IR. This is in all ways similar to the DAM case. The convergence of the integral featured in the energy estimate of Eq.~\eqref{eq:EN-ss} ensures that the wave speed $c=1/(g-3/2)$ prescribed by Eq.~\eqref{eq:ccons} remains valid in the NLS case. This ultimately prescribes the right front evolution $\propto t^{1/3}$ as previously seen in Fig.~\ref{fig:2}. The discrepancies between the numerics and the theoretical predictions seen in the right panel of Fig.~\ref{fig:7} reflect the fact that due to the presence of nonlocal effects and dissipation, the propagation $t^{1/3}$ is seen only at the level of the RJ front $\hat  \omega$ in NLS.


\subsection{Bulk}
\begin{figure}
\includegraphics[width=\textwidth]{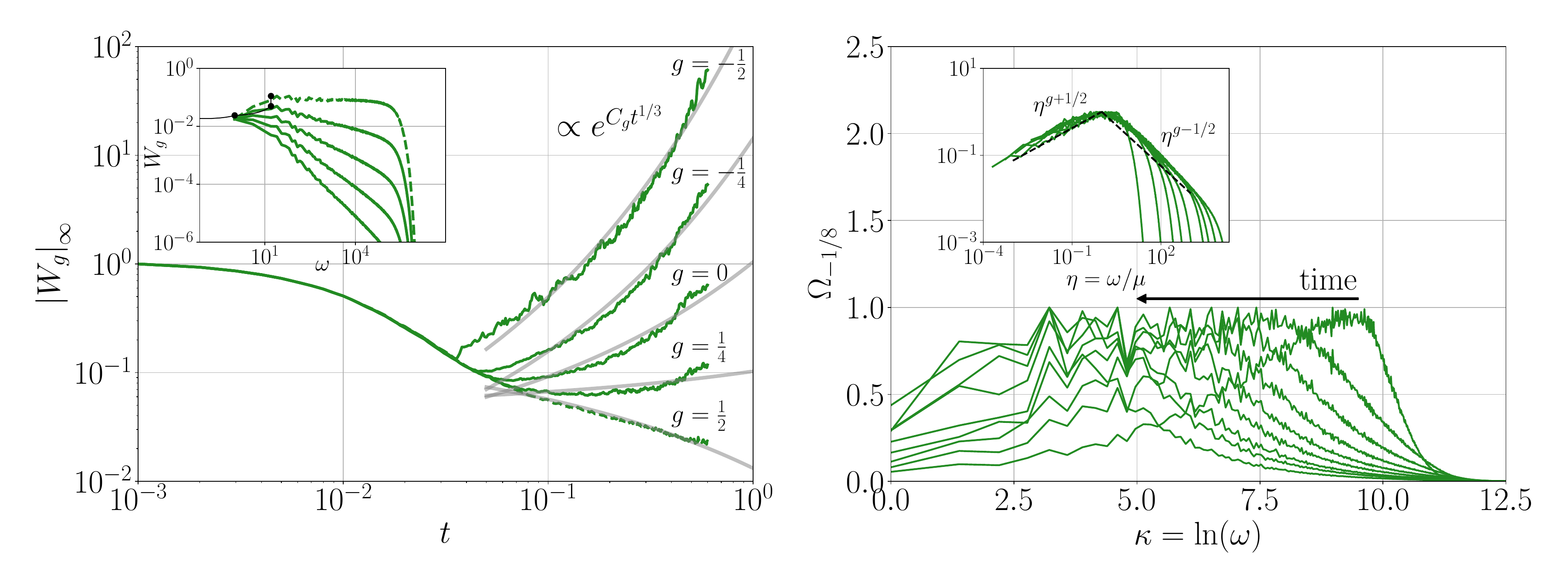}
\caption{Same as in Fig.\ref{fig:6}, but for NLS.}
\label{fig:8}
\end{figure}

Similarly to Fig.~\ref{fig:5} for the DAM, the left panel of Fig.~\ref{fig:8} presents evolution of the supremum norms of $W_g$ for $g\ge  -1/2$ obtained in the NLS simulation. Now we can see convincing numerical evidence of the self-similar evolution for the bulk part of the spectrum. The norms follow the theoretical prediction $e^{C_g t^{1/3}}$ which describes a blowup if $g\le 1/4$. The right panel of Fig.~\ref{fig:8} presents the case $g=-1/8$, showing in the inset that the profiles of $G_g$ do approach the RJ asymptotics in the central part, and the main part that (although very noisy) the profile of $\Omega_g$ does propagate as a wave to the left of the $\ln\left(\omega\right)$ axis.



\section{Discussion}
\label{sec:concluding}
\noindent In this paper, we have considered evolution of the spectrum, which is initially concentrated in a narrow frequency band described by the 2D NLS model and associated with it WKE and DAM. As such, this is an example of the classical ``ultraviolet catastrophe" phenomenon realized in a system of classical waves, where the ultimate equilibrium state has zero temperature, i.e., the energy is spread over the infinite frequency space so that the spectrum at each fixed frequency is vanishing. 
Our analysis revealed that the cooling (thermalization to zero temperature) involves parts of the spectrum exhibit different kinds of self-similar behavior. 
In DAM, there is a partition of the spectrum into IR, central, and UV regions. The central (bulk) region is quasi-thermalized: its spectrum is close to the RJ distribution with time-dependent temperature and chemical potential. The IR and UV parts terminate at a sharp left- and right-propagating fronts respectively. The process is non-uniform: each part evolves self-similarly with distinct rates and kinds. They produce divergent (blowup) behaviors for suprema of the weighted profiles $W_g  =\omega^{g+1/2}N_\omega$, which are either algebraic or stretched-exponential. The UV part exhibits a first-kind self-similar blowup directly prescribed by the conservation of energy and algebraically divergent behavior. The bulk exhibits a self-similarity, associated with stretched exponential divergent behavior. This type of self-similarity is of the third-kind, i.e., its similarity properties are determined by matching to another self-similar region, in our case the UV part. The IR part is also self-similar of the third-kind. In fact, the IR front scaling $\omega_-(t)$ can be easily obtained from the assumption that the IR and the central portions are parts of the same self-similar solution; this leads to $\omega_-(t) \propto \mu(t)$. However, the central part somewhat deviates from the RJ shape (otherwise the central profile would be stationary), and this leads to the fact that the above relation holds only in a logarithmic sense, $\ln\left(\omega_-(t)\right) \propto \ln\left(\mu(t)\right)$, as observed in the numerical simulations of the DAM. 
Overall, the DAM spectrum is characterized by universal tails with sharp fronts at both spectral ends, with the tails connected to the central part by matching regions where RJ asymptotics are approximately recovered.

The 2D NLS numerical results are mostly similar to those of the DAM except for two important differences. First, there is no left-propagating front--the spectrum quickly forms an IR plateau extending all the way to zero frequency. We explain this behavior by nonlocal WKE dynamics. Second, in the NLS simulations, the right-propagating front is affected by the finite resolution and associated hyper-viscous dissipation. As a result, numerical simulations do not validate the theoretical predictions for $W_g$ for large $g$ corresponding to the right front. However, the predicted self-similarity of the right front is indeed observed in the NLS simulations when this front is defined using a rather high threshold value. In terms of bulk dynamics, the NLS simulations show convincing evidence of the predicted stretched-exponential behavior of $W_g$ for small $g$.

\bibliographystyle{plain}
\bibliography{biblio}
\appendix

\section{Collision integral convergence}\label{app:convergence}

\noindent In this appendix, we will study the convergence of the collision integral in the frequency representation of the kinetic wave equation~\eqref{eq:kinetic_omega}. Considering a power-law spectrum ansatz of the form $N_\omega = C\omega^{-x}$, the collision integral $St[N_\omega=C\omega^{-x}]=St[x]$ simplifies to
\begin{align}\label{eq:kinetic_x}
St[x]&=C^3\int S^{\omega,1}_{2,3}(\omega\omega_1\omega_2\omega_3)^{-x}\left(\omega^x + \omega_1^x -\omega_2^x-\omega_3^x\right)\delta\left(\omega+\omega_1-\omega_2-\omega_3\right)\, \de\omega_1\,\de\omega_2\,\de\omega_3.
\end{align}
Before studying the convergence of~\eqref{eq:kinetic_x}, we first examine the behavior of the interaction kernel $S^{\omega, 1}_{2,3}$ with respect to the limiting values of the frequencies.

\subsection{Scaling behaviour of the interaction kernel $S^{\omega, 1}_{2,3}$}
\noindent 
First of all, let us recall the explicit expression for the interaction kernel as found in \cite{dyachenko_optical_1992}. It reads
\be
S_{\omega,\omega_1}^{\omega_2,\omega_3}=S({\omega},{\omega}_1,{\omega}_2,{\omega}_3) = S_1K(q)
\ee
 where $S_1 =4/\pi(\sqrt{\omega\omega_1} +\sqrt{\omega_2\omega_3})$ and $K(q)$ is the complete elliptic integral of the first-kind with elliptic modulus $q=2\sqrt[4]{\omega\omega_1\omega_2\omega_3}/(\sqrt{\omega\omega_1}+\sqrt{\omega_2\omega_3})$ \cite{dyachenko_optical_1992}. We can check that the kernel emits a degree of homogeneity of $-1$ in frequencies:  $S(\lambda{\omega},\lambda{\omega}_1,\lambda{\omega}_2,\lambda{\omega}_3) = \lambda^{-1}S({\omega},{\omega}_1,{\omega}_2,{\omega}_3)$.
 The behaviour of the complete elliptic integral can be ascertained by the Taylor expansion for small $q^2$, leading to
\begin{align*}
K(q) &= \frac{\pi}{2}\left[ 1 + \left(\frac{1}{2}\right)^2q^2 +\left(\frac{1\times 3}{2\times 4}\right)^2q^4 + \cdots+\left(\frac{(2n-1)!!}{(2n)!!}\right)^2q^{2n}+\cdots \right] \quad \text{for $q^2<1$},
\end{align*}
where $n!!$ denotes the double factorial operation. Close to the singularity at $q=1$, we can expand in the complementary modulus variable $q'=\sqrt{1-q^2}$ leading to 
\begin{align*}
K(q) &= \ln\left(\frac{4}{q'}\right) + \left(\frac{1}{2}\right)^{2}\left[\ln\left(\frac{4}{q'}\right) - \frac{2}{1\times 2}\right]{q'}^2+\left(\frac{1\times 3}{2\times 4}\right)^{2}\left[\ln\left(\frac{4}{q'}\right) - \frac{2}{1\times 2}- \frac{2}{3\times 4}\right]{q'}^4 + \dots 
\end{align*}
for $q\to1$. Therefore, we observe a logarithmic singularity as $q\to 1$, i.e., for $\omega_2\to \omega$ ($\omega_3\to \omega$) and $\omega_3\to \omega_1$ ($\omega_2\to \omega_1$). It follows, that $K(q)$ has an integrable logarithmic singularity at $q=1$.

\subsection{(a) Limit $\omega_1, \omega_2,\omega_3 \to \omega$}
\noindent Consider the limit $\omega_1, \omega_2,\omega_3 \to \omega$, which does not contradict the Dirac delta function of frequencies leading to the resonance condition $\omega+\omega_1 = \omega_2 + \omega_3$. In this limit, we observe that $q\to 1$ and $S^{\omega,1}_{2,3}$ will have a logarithmic singularity. By applying the resonance condition, $St[x]$ becomes  
\begin{align}\label{eq:1}
St[x]&=C^3\int S^{\omega,1}_{2,\omega + 1-2}\left[\omega\omega_1\omega_2(\omega+\omega_1-\omega_2)\right]^{-x}\left(\omega^x + \omega_1^x -\omega_2^x-(\omega+\omega_1-\omega_2)^x\right)\, \de\omega_1\,\de\omega_2.
\end{align}
Then, by introducing new variables $\tilde{\omega}_1 = \omega_1-\omega$ and $\tilde{\omega}_2 = \omega_2-\omega$, and using polar coordinates $\tilde{\omega}_1 = r\cos(\theta)$, $\tilde{\omega}_2 = r\sin(\theta)$ we find that
\begin{align*}
q &= \frac{2\sqrt[4]{\omega(\tilde{\omega}_1 + \omega)(\tilde{\omega}_2 + \omega)(\omega+\tilde{\omega}_1 - \tilde{\omega_2})}}{\sqrt{\omega(\tilde{\omega}_1+\omega)}+\sqrt{(\tilde{\omega}_2+\omega)(\omega+\tilde{\omega}_1 -\tilde{\omega}_2)}}\propto 1 + ar^4 + O(r^5), \quad \text{as $r\to 0$},
\end{align*}
where $a = -\sin^2(\theta)(\cos(\theta)-\sin(\theta))^2/(32\omega^4)$. Subsequently, $S^{\omega,1}_{2,\omega+1-2}$ contains a logarithmic singularity that grows as $S^{\omega,1}_{2,\omega+1-2}\propto \ln(1/q') \propto \ln(r^2)$ as $r\to 0$. The first square bracket term has a scaling as $r\to 0$ of
\begin{align*}
\left[\omega(\tilde{\omega}_1+\omega)(\tilde{\omega}_2+\omega)(\omega+\tilde{\omega}_1-\tilde{\omega}_2)\right]^{-x} \propto \omega^{-4x} + O(r),
\end{align*}
while the second square bracket term in Eq~\eqref{eq:1} yields, at leading order
\begin{align*}
&\omega^x + \omega_1^x -\omega_2^x-(\omega+\omega_1-\omega_2)^x= \omega^x + (\tilde{\omega}_1+\omega)^x -(\tilde{\omega}_2+\omega)^x-(\omega+\tilde{\omega}_1-\tilde{\omega}_2)^x \propto r^2. 
\end{align*}
Therefore, with the additional change of variables element of $\de\omega_1\, \de\omega_2 = r\, \de r\, \de \theta$, we find that the collision integral becomes regularised  in the $r\to 0$ limit, for any power of $x$, with
\begin{align*}
St[x] \propto \int_0 r^3\ln(r^2)\, \de r < \infty.
\end{align*}

\subsection{(b) Limit $\omega_1 \ll \omega_2 \sim \omega_3 \sim \omega$}

\noindent Considering Eq.~\eqref{eq:1} in the limit of $\omega_1\to 0$, while $\omega_2,\omega_3 \sim \omega \sim 1$. With this assumption $q\propto \omega_1^{1/4}$ which implies that $q^2\ll 1$, and hence $K(q)\to \pi/2$, and $S^{\omega,1}_{2,\omega+1-2}\propto 1$. In addition, we find that $\left[\omega\omega_1\omega_2(\omega+\omega_1-\omega_2)\right]^{-x} \propto \omega_1^{-x}$, while the second squared bracket term in Eq.~\eqref{eq:1} yields at leading order
\begin{align*}
\omega^x + \omega_1^x -\omega_2^x-(\omega+\omega_1-\omega_2)^x \propto \begin{cases} 1 & \text{for $x>0$,}\\ \omega_1^{x} & \text{for $x<0$}. \end{cases}
\end{align*}
Therefore,
\begin{align*}
St[x] \propto \begin{cases} \int_0 \omega_1^{-x}\, \de\omega_1  & \text{for $x>0$}\\ \int_0 \omega_1^{0}\, \de\omega_1 < \infty & \text{for $x<0$}, \end{cases}
\end{align*}
and therefore is convergent for $x<1$.

\subsection{(c) Limit $\omega_1, \omega_2 \ll \omega_3 \sim \omega$}
\noindent Taking into account the limit $\omega_1, \omega_2 \ll 1$, and $\omega_3 \sim \omega\sim 1$, consider the polar coordinate decomposition $\omega_1 = r\cos(\theta)$ and $\omega_2=r\sin(\theta)$. Analysing the behaviour of $q$, we find that $q<1$ and therefore $K(q)<\infty$. Consequently, we find $S^{\omega,1}_{2,\omega+1-2} \propto r^{-1/2}$. The first square bracket in~\eqref{eq:1} leads to $\left[\omega\omega_1\omega_2(\omega+\omega_1-\omega_2)\right]^{-x} \propto r^{-2x}$, while the second square bracket term provides a leading behaviour of
\begin{align*}
\omega^x + \omega_1^x -\omega_2^x-(\omega+\omega_1-\omega_2)^x \propto \begin{cases} r^{x} & \text{for $x<1$,}\\ r & \text{for $1\leq x$}. \end{cases} 
\end{align*}
Therefore, we find that
\begin{align*}
St[x] \propto \begin{cases} \int_0 r^{(1-2x)/2}\, \de r & \text{for $x < 1$}, \\  \int_0 r^{(3-4x)/2}\, \de r  & \text{for $1\leq x$}.\end{cases} 
\end{align*}
which collectively implies that the collision integral $St[x]$ converges in this limit as long as $x< 5/4$.

\subsection{(d) Limit $\omega \ll \omega_1, \omega_2 , \omega_3$}
\noindent Consider three frequencies $1 \ll \omega_1, \omega_2, \omega_3$, while $ \omega\sim 1$.  We use polar coordinates $\omega_1=r\cos(\theta)$ and $\omega_2=r\sin(\theta)$ and consider $\theta\neq \pi/4$. The latter condition leads to large $\omega_3=\omega+\omega_1-\omega_2$ as $r\to \infty$. Subsequently, we find that $q \propto r^{-1/4}$ as $r\to \infty$, which implies $q\ll 1$, and $K(q)\to \pi/2$. Consequently, we find $S^{\omega,1}_{2,\omega+1-2} \propto r^{-1}$. The first square bracket term in Eq.~\eqref{eq:1} leads to $\left[\omega\omega_1\omega_2(\omega+\omega_1-\omega_2)\right]^{-x} \propto  r^{-3x}$, while the second square bracket term provides
\begin{align*}
\omega^x + \omega_1^x -\omega_2^x-(\omega+\omega_1-\omega_2)^x \propto \begin{cases} r^{0} & \text{for $x<0$,}\\ r^{x} & \text{for $0\leq x$}.\end{cases} 
\end{align*}
Therefore, we find that in this limit
\begin{align*}
St[x] \propto \begin{cases} \int^\infty r^{-3x}\, \de r & \text{for $x < 0$} \\ \int^\infty r^{-2x}\, \de r  & \text{for $0\leq x$}\end{cases} 
\end{align*}
which implies convergence for $1/2<x$.

\subsection{(e) Limit $\omega_3 \sim \omega\ll \omega_1, \omega_2$}
\noindent Considering two large frequencies $1 \ll \omega_1, \omega_2$, while $\omega_3 \sim \omega\sim 1$, we study a special case of subsection (d) where $\theta \to \pi/4$. In such a situation, we observe that $q\to 1$ meaning that $S^{\omega,1}_{2,\omega+1-2}$ contains a logarithmic singularity, but is regularised by the scaling of $S_1$, leading to $S^{\omega,1}_{2,\omega+1-2}\propto r^{-1/2}$ The first square bracket term in~\eqref{eq:1} leads to $\left[\omega\omega_1\omega_2(\omega+\omega_1-\omega_2)\right]^{-x} \propto  r^{-2x}$, while the second square bracket term $\omega^x + \omega_1^x -\omega_2^x-(\omega+\omega_1-\omega_2)^x=0$ completely cancels leading to the collision integral vanishing for any spectral exponent $x$ in this special case.

\subsection{Summary on the convergence criterion of the collision integral $St[x]$.}

\begin{figure*}[htp!]
\centering
\scalebox{0.81}{
\begin{tikzpicture}[domain=-0.12:1.5,scale=10] 
	\draw(-0.12,0.125) node[anchor=east, inner sep = 17pt]{Convergence Region};
	\draw[fill=green!50!black!50, green!50!black!50] (0.5,0.1) rectangle (1.0,0.15);
	\draw[fill=red!50!black!50, red!50!black!50] (-0.1,0.1) rectangle (0.5,0.15);
	\draw[fill=red!50!black!50, red!50!black!50] (1,0.1) rectangle (1.5,0.15);

    \draw[fill=green!50!black!50, green!50!black!50] (-0.005,0.1) rectangle (0.005,0.15);
    \draw[fill=green!50!black!50, green!50!black!50] (0.995,0.1) rectangle (1.005,0.15);

	\draw(-0.16,-0.025) node[anchor=east, inner sep = 16pt]{(a) $  \omega_1,\omega_2,\omega_3\to\omega$};
	\draw[fill=green!50!black!50, green!50!black!50] (-0.1,0) rectangle (1.5,-0.05);
	
\draw(-0.14,-0.075) node[anchor=east]{(b) $\omega_1\ll\omega_2\sim\omega_3\sim\omega$};
	\draw[fill=green!50!black!50, green!50!black!50] (-0.1,-0.05) rectangle (1.0,-0.1);
	\draw[fill=red!50!black!50, red!50!black!50] (1.0,-0.05) rectangle (1.5,-0.1);

	\draw(-0.15,-0.125) node[anchor=east, inner sep = 12pt]{(c) $\omega_1, \omega_2\ll \omega_3\sim\omega$};
	\draw[fill=green!50!black!50, green!50!black!50] (-0.1,-0.1) rectangle (1.25,-0.15);
	\draw[fill=red!50!black!50, red!50!black!50] (1.25,-0.1) rectangle (1.5,-0.15);

	\draw(-0.06,-0.175) node[anchor=east, inner sep = 28pt]{(d) $\omega \ll \omega_1 \sim \omega_2 \sim \omega_3$};
	\draw[fill=red!50!black!50, red!50!black!50] (-0.1,-0.15) rectangle (0.5,-0.2);
	\draw[fill=green!50!black!50, green!50!black!50] (0.5,-0.15) rectangle (1.5,-0.2);

	\draw(-0.12,-0.225) node[anchor=east, inner sep = 12pt]{(e) $\omega \sim \omega_3 \ll \omega_1 \sim \omega_2$};
	\draw[fill=green!50!black!50, green!50!black!50] (-0.1,-0.2) rectangle (1.5,-0.25);

	\draw(0,0.05) node{$0$};
	\draw(1/2,0.05) node{$1/2$};
	\draw(5/4,0.05) node{$5/4$};
	\draw(1,0.05) node{$1$};
	\draw[-] (0,0.02) -- (0,-0.25);
	\draw[-] (1/2,0.02) -- (1/2,-0.25);
	\draw[-] (1.0,0.02) -- (1.0,-0.25);
	\draw[-] (5/4,0.02) -- (5/4,-0.25);

	\draw[-] (0,0.08) -- (0,0.15);
	\draw[-] (1/2,0.08) -- (1/2, 0.15);
	\draw[-] (1.0,0.08) -- (1.0, 0.15);
	\draw[-] (5/4,0.08) -- (5/4, 0.15);

	\draw[->,line width=2pt] (-0.1,0) -- (1.52,0.0) node[right] {$x$};

	\draw[dashed,line width=1pt,gray] (-0.1,-0.05) -- (1.5,-0.05); 
	\draw[dashed,line width=1pt,gray] (-0.1,-0.1) -- (1.5,-0.1); 
	\draw[dashed,line width=1pt,gray] (-0.1,-0.15) -- (1.5,-0.15); 
	\draw[dashed,line width=1pt,gray] (-0.1,-0.2) -- (1.5,-0.2); 

	\draw(1.25,-0.075) node{divergent};
	\draw(1.375,-0.125) node{divergent};
	\draw(0.25,-0.175) node{divergent};

		\draw(0.25,0.125) node{divergent};	
    	\draw(0.75,0.125) node{convergent};	
      	\draw(1.125,0.125) node{divergent};	
\end{tikzpicture}
}
\caption{Depiction of the convergence (green) regions of the WKE~\eqref{eq:kinetic_omega} for power-law spectral slope $x$, where $N_\omega = C\omega^{-x}$, in various limiting regions of frequency space as outlined in the far left column. The overall convergence criterion is presented in the top horizontal bar.\label{fig:convergence}}
\end{figure*}
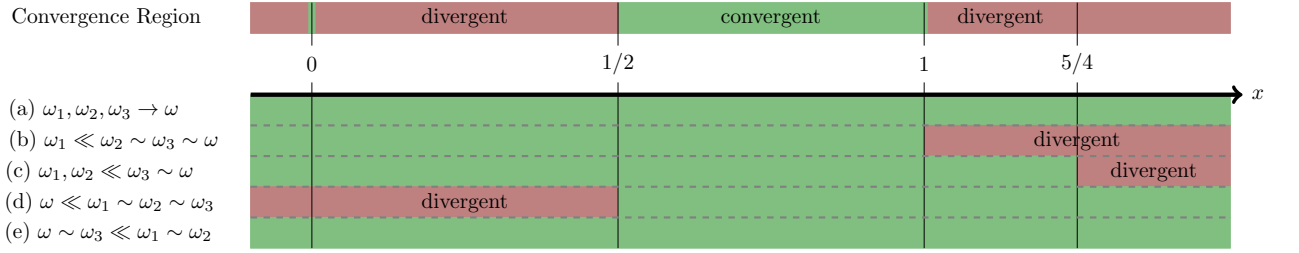

To summarise subsections (a)-(e), we display the convergence criteria in Fig.~\ref{fig:convergence} using green for convergent and red of divergent. Consequently, we find that the collision integral is convergent for $1/2<x\leq 1$. 

\section{Analysis of the Nonlocal Regions}
\label{app:nonlocal}

\subsection{Nonlocality at low frequencies - for $\omega \ll \omega_1, \omega_2, \omega_3$}

The system evolves toward a state close to the Rayleigh–Jeans distribution at low frequencies, as discussed in the main text. Consequently, the low-frequency power-law behaviour approaches a constant. According to Appendix~\ref{app:convergence}, case (e), this implies nonlocal behaviour; that is, the spectral evolution at low $\omega$ is dominated by interactions with much higher frequencies. The purpose of this Appendix is to examine a self-similar evolution of the spectrum within this regime.

Consider the dominate nonlocal interaction from case (e) $\omega_1, \omega_2, \omega_3 \gg \omega$. In this case we have $q\ll 1$, $K(q)=\pi/2$, and $S_1=4/\pi\sqrt{\omega_2\omega_3}$. Then
\begin{align}\label{eq:kinetic_nonlocal}
\frac{\de N_\omega}{\de t}&= \int \frac{1}{2\sqrt{\omega_2\omega_3}} N_1N_2 N_3\delta(\omega_1-\omega_2-\omega_3)\ \de\omega_1\,\de\omega_2\,\de\omega_3\nonumber \\
& + N_\omega\int  \frac{1}{2\sqrt{\omega_2\omega_3}}N_1N_2 N_3\left(\frac{1}{N_1}-\frac{1}{N_2}-\frac{1}{N_3}\right)\delta(\omega_1-\omega_2-\omega_3)\ \de\omega_1\,\de\omega_2\, \de\omega_3.
\end{align}
Denote the first integral in~\eqref{eq:kinetic_nonlocal} by $A(t)$ and the second by $B(t)$ we can express~\eqref{eq:kinetic_nonlocal} as
\begin{align}\label{eq:selfsimilar_system}
\frac{\de N_\omega}{\de t}&= A(t) + B(t)N_\omega.
\end{align}
The integrals $A(t)$ and $B(t)$ on the right-hand side no longer depend on $\omega$ and $N_\omega$. Looking for a self-similar solution of~\eqref{eq:selfsimilar_system} in the form $N_\omega = t^a f(\eta)$ where $\eta =\omega/t^{b}$ such that $b=-2a-1 < 0$. Then Eq.~\eqref{eq:selfsimilar_system} reads
\begin{align}\label{eq:selfsimilar_system2}
{a}f-b\,\eta \frac{\de f}{\de\eta} = 
{\tilde{A}} + {\tilde{B}}f
\end{align}
where $\tilde{A}$ and $\tilde{B}$ represent the self-similar counterpart of the integrals $A(t)$ and $B(t)$, given by
\begin{align}
\tilde{A}  &= \int  \frac{1}{2\sqrt{\eta_2\eta_3}}f_1f_2 f_3\delta(\eta_1-\eta_2-\eta_3)\ \de\eta_1\,\de\eta_2\, \de\eta_3\\
\tilde{B}  &= \int  \frac{1}{2\sqrt{\eta_2\eta_3}}f_1f_2 f_3\left(\frac{1}{f_1}-\frac{1}{f_2}-\frac{1}{f_3}\right)\delta(\eta_1-\eta_2-\eta_3)\ \de\eta_1\,\de\eta_2\, \de\eta_3.
\end{align}
It is easy to see that $\tilde{A}$ and $\tilde{B}$ are independent of $\eta$. Equation~\eqref{eq:selfsimilar_system2} can be easily integrated to give a general solution 
\begin{align}
f(\eta) = \frac{\tilde{A}}{a-\tilde{B}} + C\eta^{(a-\tilde{B})/b}.
\end{align}
Since $f,\tilde A>0$, we must have $a>\tilde B$ -- else $f(\eta)$ is either a negative constant of it changes sign at some $\eta$. Then the exponent in the second term is negative (because $b<0$) and the only way to avoid sinfularity at $\eta \to 0$ is to put 
$C=0$, i.e.
\begin{align}\label{eq:selfsimilar-sol}
f(\eta) = \frac{\tilde{A}}{a-\tilde{B}}.
\end{align}
Result~\eqref{eq:selfsimilar-sol} implies that the only self-similar behaviour governed by nonlocal interaction in the low-frequency region is for the spectrum to have a constant value, evolving according to the behaviours of $\tilde{A}$ and $\tilde{B}$. Ultimately, as the DAM does not take into account the possibility of nonlocality (inherent from the DAM superlocal interaction approximation), the full 2D NLS evolution will differ to the DAM in the low frequency region. Indeed, we observe such plateau characteristic of the wave action spectrum for the 2D NLS in Fig.~\ref{fig:1} (right), and the lack of a  low-frequency left front $\omega_-$.

\subsection{Nonlocality at high frequencies}


At high frequencies, we have a nonlocality of the collision integral occuring for case (b) in Appendix~\ref{app:convergence} for a spectrum steeper than $N_\omega \propto 1/\omega$. In this case we should consider dominance of contributions to the collision integral from the nonlocal interactions with the IR end, i.e., where one frequency $\omega_1 \ll \omega$ and $\omega_2,\omega_3\sim \omega$ (and similarly for the $\omega_2\ll \omega$ and $\omega_3 \ll \omega$ cases). Then by the definition of the elliptic modulus $q\ll 1$ and hence $K(q)=\pi/2$. With this the kinetic equation~\eqref{eq:kinetic_omega} can be written in the form
\begin{align*}
\frac{\de N_\omega}{\de t} \approx \int \frac{1}{2\left(\sqrt{\omega\omega_1} + \sqrt{\omega_2\omega_3}\right)}N_\omega N_1N_2 N_3\left(\frac{1}{N_\omega}+\frac{1}{N_1}-\frac{1}{N_2}-\frac{1}{N_3}\right)\delta(\omega+\omega_1-\omega_2-\omega_3)\ \de\omega_1\,\de\omega_2\,\de\omega_3.
\end{align*}
Consider the integral which is responsible for leading contribution when $\omega_1 \ll \omega$, then the leading contribution (denoted as $I_1$) becomes
\begin{align*}
I_1= \frac{1}{2} \int \frac{1}{\sqrt{\omega_2\omega_3}}N_\omega N_1N_2 N_3\left(\frac{1}{N_\omega}-\frac{1}{N_2}-\frac{1}{N_3}\right)\delta(\omega-\omega_2-\omega_3)\ \de\omega_1\,\de\omega_2\,\de\omega_3,
\end{align*}
and integrating over $\omega_1$ we get
\begin{align*}
I_1&= \frac{1}{2}\int N_1\, \de\omega_1 \int \frac{1}{\sqrt{\omega_2\omega_3}}N_\omega N_2 N_3\left(\frac{1}{N_\omega}-\frac{1}{N_2}-\frac{1}{N_3}\right)\delta(\omega-\omega_2-\omega_3)\ \de\omega_2\,\de\omega_3\\
&= C(t) \int \frac{1}{\sqrt{\omega_2\omega_3}}N_\omega N_2 N_3\left(\frac{1}{N_\omega}-\frac{1}{N_2}-\frac{1}{N_3}\right)\delta(\omega-\omega_2-\omega_3)\ \de\omega_2\,\de\omega_3,
\end{align*}
where $C(t)= (1/2)\int N_1\, \de\omega_1$. Using the notation from~\cite{nazarenko_wave_2011} we can write
\begin{align*}
{\rm St}[N_\omega] \approx I_1= C(t)\int R^\omega_{2,3} \ \de\omega_2\, \de\omega_3,
\end{align*}
where $R^{\omega}_{2,3}=(1/\sqrt{\omega_2\omega_3})N_\omega N_2 N_3\left(\frac{1}{N_\omega}-\frac{1}{N_2}-\frac{1}{N_3}\right)\delta(\omega-\omega_2-\omega_3)$. The two remaining integrals $I_2$ and $I_3$ in the respective limits $\omega_2\ll 1$ and $\omega_3\ll 1$  both lead to $S_1\approx 4/\pi\sqrt{\omega\omega_1}$ add further contributions to the nonlocality at high frequencies. Subsequently,
\begin{align*}
I_2&= \frac{1}{2} \int \frac{1}{\sqrt{\omega\omega_1}}N_\omega N_1N_2 N_3\left(\frac{1}{N_\omega}+\frac{1}{N_1}-\frac{1}{N_3}\right)\delta(\omega+\omega_1-\omega_3)\ \de\omega_1\,\de\omega_2\,\de\omega_3\\
&= C(t)\int \frac{1}{\sqrt{\omega\omega_1}}N_\omega N_1 N_3\left(\frac{1}{N_\omega}+\frac{1}{N_1}-\frac{1}{N_3}\right)\delta(\omega+\omega_1-\omega_3)\ \de\omega_1\,\de\omega_3\\
&=-C(t)\int R^3_{2,\omega} \ \de\omega_2\, \de\omega_3.
\end{align*}
and 
\begin{align*}
I_3&= \frac{1}{2} \int \frac{1}{\sqrt{\omega\omega_1}}N_\omega N_1N_2 N_3\left(\frac{1}{N_\omega}+\frac{1}{N_1}-\frac{1}{N_2}\right)\delta(\omega+\omega_1-\omega_2)\ \de\omega_1\,\de\omega_2\,\de\omega_3\\
&= C(t)\int \frac{1}{\sqrt{\omega\omega_1}}N_\omega N_1 N_3\left(\frac{1}{N_\omega}+\frac{1}{N_1}-\frac{1}{N_2}\right)\delta(\omega+\omega_1-\omega_2)\ \de\omega_1\,\de\omega_2\\
&=-C(t)\int R^2_{3,\omega} \ \de\omega_2\, \de\omega_3.
\end{align*}
Hence, we get a standard three-wave kinetic equation that describes the leading nonlocal interaction at large frequencies:
\begin{align*}
\frac{\de N_\omega}{\de t}\approx I_1 + I_2 + I_3 = C(t)\int \left(R^\omega_{2,3}-R^3_{2,\omega}- R^2_{3,\omega}\right) \, \de\omega_2\, \de \omega_3.
\end{align*}
Based on this equation, we can explain why solutions of the WKE cannot have a sharp right front (falling to zero at a finite frequency). Indeed, let a an initial spectrum have a sharp front $\omega_+$, and let us consider a frequency at a finite, but not too large, distance to the right of the front, $\omega -\omega_+ <\sigma$, where $\sigma$ is the width of the initial support of the spectrum. Then there exist a resonant triads $\omega, \omega_2, \omega_3, \omega=\omega_2+\omega_3$, such that $\omega_2$ and $\omega_3$ are within the initial support. Therefore, at any arbitrarily small time the spectrum at $\omega$ would become nonzero, which would imply an infinite speed of $\omega_+(t)$. In  other words, 
at any positive time, the spectrum support will be infinite, without a finite boundary 
$\omega_+$.

\end{document}